**Title**
mHealth hyperspectral learning for instantaneous spatiospectral imaging of hemodynamics

**Authors**
Yuhyun Ji[1], Sang Mok Park[1], Semin Kwon[1], Jung Woo Leem[1], Vidhya Vijayakrishnan Nair[1], Yunjie Tong[1], and Young L. Kim[1,2,3,4]*
[1]Weldon School of Biomedical Engineering, Purdue University, West Lafayette, IN, USA
[2]Purdue Institue for Cancer Research, West Lafayette, IN, USA
[3]Regenstrief Center for Healthcare Engineering, West Lafayette, IN, USA
[4]Purdue Quantum Science and Engineering Institute, West Lafayette, IN, USA
* youngkim@purdue.edu

**Abstract**
Hyperspectral imaging acquires data in both the spatial and frequency domains to offer abundant physical or biological information. However, conventional hyperspectral imaging has intrinsic limitations of bulky instruments, slow data acquisition rate, and spatiospectral tradeoff. Here we introduce hyperspectral learning for snapshot hyperspectral imaging in which sampled hyperspectral data in a small subarea are incorporated into a learning algorithm to recover the hypercube. Hyperspectral learning exploits the idea that a photograph is more than merely a picture and contains detailed spectral information. A small sampling of hyperspectral data enables spectrally informed learning to recover a hypercube from a red-green-blue (RGB) image without complete hyperspectral measurements. Hyperspectral learning is capable of recovering full spectroscopic resolution in the hypercube, comparable to high spectral resolutions of scientific spectrometers. Hyperspectral learning also enables ultrafast dynamic imaging, leveraging ultraslow video recording in an off-the-shelf smartphone, given that a video comprises a time series of multiple RGB images. To demonstrate its versatility, an experimental model of vascular development is used to extract hemodynamic parameters via statistical and deep-learning approaches. Subsequently, the hemodynamics of peripheral microcirculation is assessed at an ultrafast temporal resolution up to a millisecond, using a conventional smartphone camera. This spectrally informed learning method is analogous to compressed sensing; however, it further allows for reliable hypercube recovery and key feature extractions with a transparent learning algorithm. This learning-powered snapshot hyperspectral imaging method yields high spectral and temporal resolutions and eliminates the spatiospectral tradeoff, offering simple hardware requirements and potential applications of various machine-learning techniques.

**Significance Statement**
The development of ultrafast hyperspectral imaging systems is an active area of research for a variety of applications from biomedical to defense domains. However, such hardware-focused approaches have been challenging mainly because of the intrinsic tradeoff between spatial and spectral resolutions and the slow rates of data acquisition. Hyperspectral learning takes inspiration from the idea that a photograph is more than merely a picture and indeed contains rich spectral information. Hyperspectral learning enables the recovery of spectral information with high spectral and temporal resolutions from RGB color values acquired using a conventional camera. The main advantages of hyperspectral learning include hardware simplicity, no tradeoff between spatial and spectral resolutions, high temporal resolution, and machine learning augmentation for snapshot imaging.

**Introduction**
Hyperspectral (with a high spectral resolution of ~10 nm) or multispectral (with several spectral bands of ~50 nm) imaging systems acquire a hyperspectral image dataset (hypercube)–a three-dimensional dataset of spectral intensity in spatial coordinates. Subsequently, both the spatial and spectral data are processed [1-3]. Hyperspectral imaging technologies offer extensive physical and biological information in stationary or dynamic samples, ranging from microscopic settings to airborne remote-sensing environments, for a variety of applications in geology, mineralogy, agriculture, environmental science, astronomy, forensic medicine, defense, security, and biomedicine [4-9]. Notably, hyperspectral imaging technologies have been reinvigorated through recent advances in data-driven machine learning [2, 10-14]. For example, deep-learning approaches have enabled the effective processing of extremely large hypercube data for classical imaging tasks [11, 15-17] and allowed for the optimization of hypercube acquisition to achieve specific tasks and objectives [18-22]. Data fusion of complementary images with high-spectral or high-spatial resolutions [23, 24] and neural networks of improving spatial resolutions [25] can overcome the intrinsic trade-off between spatial and spectral resolutions. However, conventional hyperspectral imaging systems still face the intrinsic limitations: bulky instruments, slow data acquisition rates, low detection efficacy (i.e., low signal-to-noise ratio), and motion artifacts [1, 2].

Typically, hyperspectral imaging systems rely on mechanical scanning elements either in the spectral or spatial domains. In particular, spectral scanning systems employ a number of narrow bandpass spectral filters or dispersive optical components, whereas point scanning and line-scanning systems rely on mechanical translational components that require high precision [1, 2, 6, 7, 9]. Thus, these scanning elements result in bulky instruments and yield suboptimal temporal resolutions. In particular, prolonged time of data acquisition time fundamentally limits dynamic imaging with a high temporal resolution. In this respect, the development of snapshot imaging technologies capable of acquiring a hypercube in a single shot manner has been an active area of research (summarized in Table S1). The most common configuration used for snapshot imaging involves simultaneously capturing multiple images with different spectral bands using a large-area image sensor [26-32]. In particular, large-area image sensor-based snapshot imaging is beneficial for reducing the acquisition time [2, 33, 34]. Other snapshot-imaging technologies employ dispersion patterns or coded apertures projecting irradiance mixed with spatial and spectral information to further enhance the light-collection efficiency and readout rate [26, 27, 35-37]. Subsequently, the modulated projection comprising spatial and spectral information is reconstructed into a hypercube by utilizing computational algorithms such as compressed (or compressive) sensing [32, 38, 39], Fourier transformation [40, 41], or deep learning [19, 20, 42-48].

However, previously developed hyperspectral imaging technologies with a snapshot ability face several limitations [1, 2, 9, 46, 49]. First, typical snapshot systems are limited by the intrinsic tradeoff that must be made between the spectral and spatial resolutions; an improvement in spatial resolution causes a deterioration in the number of spectral bands, thereby compromising the spectral resolution or the spatial resolution (or imaging area). Second, snapshot imaging systems are sensitive to light conditions and imaging configurations, thereby introducing significant errors in field applications. Third, the hyperspectral filter arrays, dispersion patterns, and coded apertures require high-precision fabrication or nanofabrication, including precision alignment of array components, optimized miniaturization, integration with pixel-level filters, and customized calibrations, all of which inhibit manufacturability. Consequently, the previous studies have generally been performed under laboratory settings [50, 51] or with stationary biological samples [7, 9, 52], thereby hampering the practical and widespread utilization.

To develop an alternative to instantaneous hyperspectral imaging, we take advantage of spectral learning (also known as spectral super-resolution), which enables the recovery of spectral information from RGB values (tristimulus) acquired by a conventional trichromatic camera; a full reflectance spectrum in the visible range is computationally reconstructed from an RGB image. Several scientific communities, especially in the fields of color science, machine vision, and biomedical imaging have addressed this ill-posed problem using different but related methods including compressed sensing, machine learning, and deep learning [26, 30, 53-67]. Owing to its hardware simplicity, spectral learning can be performed by using a smartphone camera. In practice, dispersive optical components, such as spectrometers and bulky optical filters, are usually attached to mobile health (mHealth) sensing applications, potentially weakening user acceptance and hampering the practical translation from research to clinical practice [68-74].

Conversely, spectral learning requires only a built-in camera, potentially offering hardware-independent mHealth tools using unmodified smartphones.

In this paper, we introduce a learning-based spatiospectral imaging method offering high spectral and temporal resolutions. The proposed spectral learning involves mapping from a sparse spectral space (i.e., RGB values) to a dense spectral space. Specifically, the spectral resolution is in a range of 0.5 – 1 nm, comparable to those of scientific spectrometers and spectrographs for biomedical or biochemical applications (thereby referred to as hyperspectral learning, compared with spectral learning). Hyperspectral learning also allows us to use a video frame rate with a high temporal resolution. First, we construct a customized dual-channel imaging setup coupled with a trichromatic camera (e.g., smartphone camera) and a spectrograph to simultaneously acquire an RGB image and subarea hyperspectral data. Second, we establish a simple statistical assumption to infer the entire field-of-view from a sampled subarea and recover a hypercube from incomplete measurements. Third, we establish two complementary machine-learning frameworks based on statistical and deep learning, incorporating the domain knowledge of tissue optics into learning algorithms. Finally, we demonstrate reliable extractions of hemodynamic parameters from several different samples of tissue phantoms, chick embryos, and human conjunctiva; the results are validated through conventional hyperspectral imaging and functional near-infrared spectroscopy. Moreover, this hyperspectral learning method is applied to smartphone video recording to demonstrate the dynamic imaging of peripheral microcirculation and ultrafast imaging of oxygen depletion in tissue phantoms.

**Results**
Figure 1A illustrates the concept of hyperspectral learning for instantaneous spatiospectral imaging by significantly minimizing the number of necessary hyperspectral measurements. If hyperspectral data in a small yet representative subarea are available, a hyperspectral learning algorithm can be trained using the RGB and hyperspectral data in the subarea. This hyperspectral learning algorithm trained by the sampled (RGB and hyperspectral) data is applied to the entire image area, generating a hypercube without the need for a complete spectral or spatial scan. Thus, the key advantages of hyperspectral learning and hypercube recovery include the hardware simplicity offered by the use of conventional cameras, high temporal resolution if a video is used (e.g., slow-motion video recording on a smartphone), independence (no tradeoff) between the spatial and spectral resolutions, and abundant spectral information for a variety of machine-learning techniques. Locally sampled hyperspectral data can serve as prior information or physical constraints for incorporating domain-specific modeling into the learning algorithm, extracting critical features and parameters, and resulting in explainable and interpretable neural networks [64, 75-79].

To instantaneously sample hyperspectral data in a small subarea, a trichromatic camera (e.g., smartphone camera) is combined with a line-scan spectrograph (Fig. 1B). This configuration is commonly used in amateur telescopes that are equipped with astronomical spectroscopy systems [80]. Specifically, a dual-channel spectrograph with a photometric slit simultaneously acquires an RGB image in the entire area and the hyperspectral data of a subarea (i.e., central line) in a single-shot manner (Fig. 1B and Methods). The field-of-view is as small as 2.5 mm × 2 mm with a spatial resolution of 55 µm (Fig. S1). The sampled hyperspectral data have a spectral range of $\lambda$ = 380 – 720 nm with a spectral resolution $\Delta\lambda$ = 0.5 nm (Fig. S2 and Methods). The dual-channel imaging setup provides sufficient training data (750 – 1500 data points). This dataset is randomly split into training (80%) and testing (20%) datasets for effectively training the hyperspectral learning algorithm to be applied to the entire area eventually. In addition, this imaging setup allows us to use a smartphone camera that can acquire videos at different frame rates. In particular, highly dynamic imaging is even possible with a high temporal resolution even up to 0.0005 sec for 1920 frames per second, using commercially available smartphone models (at the time of publication).

Hyperspectral learning addresses an ill-posed problem, which is also known as spectral super-resolution and hyperspectral reconstruction. The mathematical relationship between the RGB and hyperspectral intensity is described:

$$\boldsymbol{x}_{3\times 1} = \boldsymbol{S}_{3\times k}\boldsymbol{y}_{k\times 1} + \boldsymbol{e}_{3\times 1}, \tag{1}$$

where $x$ denotes a 3 × 1 vector corresponding to three color values in the R, G, and B channels ($\boldsymbol{x} = [R, G, B]^{\mathrm{T}}$), $\boldsymbol{S}$ represents a 3 × $k$ matrix of the RGB spectral response of the three-color sensor the spectral response functions in the R, G, and B channels of the smartphone camera (also known as the sensitivity function of the camera) [81, 82], $\boldsymbol{y}$ is a $k$ × 1 vector that has the spectral intensity ($\boldsymbol{y} = [I(\lambda_1), I(\lambda_2), ..., I(\lambda_k)]^{\mathrm{T}}$) where λ is discretized in the visible range with a spectral interval of 1 nm (Methods), and $\boldsymbol{e}$ symbolizes a 3 × 1 vector of the system noise. Basically, hyperspectral learning is to obtain a pseudoinverse of $\boldsymbol{S}_{3\times k}$. Depending on the availability of training data and the desired spectral resolution, three machine-learning approaches can be used for solving this underdetermined problem: $l_1$-norm minimization (i.e., compressed sensing or sparsity regularization), $l_2$-norm minimization (i.e., least-squares regression), and deep learning [26, 30, 53-65]. Among these approaches, statistical learning using fixed-design linear regression with polynomial expansions offers a highly stable inverse calculation that can transform RGB data to high-resolution spectral data owing to the nature of $l_2$-norm minimization (Methods) [55, 62, 81, 83, 84].

The key assumption for reliable hyperspectral learning is that a sampling distribution (i.e., RGB values of the sampled subarea) should follow the parent distribution (i.e., RGB values of the entire image area); the intensity distributions between the sampled subarea and the entire field-of-view of interest are statistically identical. Specifically, the probability distribution of the R, G, and B values in the subarea needs to conform to those in the entire area in terms of variability and shape. In addition, to reliably predict unknown hyperspectral output responses from RGB values outside the subarea, the hyperspectral learning algorithm should be applied within the same (minimum and maximum) range of sampled RGB values used to train the algorithm. In a similar manner to nonparametric tests with non-Gaussian distributions, quantile-quantile (Q-Q) plots can conveniently be used to assess if the two sets of data plausibly follow the same distribution within the same range. This assumption of interpolation offers an important advantage over conventional snapshot hyperspectral imaging. If these assumptions are valid, then the hyperspectral learning is not limited by the intrinsic tradeoff between spatial and spectral resolutions.

Spectrally informed learning allows for the incorporation of physical and biological understanding of domain knowledge into learning algorithms. Typically, purely data-driven learning requires a large volume of training data and lacks explainable and interpretable learning [76, 78]. Among the various snapshot imaging applications, we focus on extracting biological parameters or spectral signatures from a hypercube using the domain knowledge of tissue optics. In this perspective, light propagation in tissue can be explained by the theory of radiative transport and robust approximations (e.g., diffusion, Born, and empirical modeling) [85-90]. Specifically, taking advantage of tissue reflectance spectral modeling (Methods), we extract the key hemodynamic parameters: oxygenated hemoglobin (HbO$_2$), deoxygenated hemoglobin (Hb), and oxygen saturation ($sPO_2$), which are the fundamental determinants of oxygen transport to tissue associated with a variety of physiological changes, diseases, and disorders.

$$sPO2 = \frac{HbO_2}{\text{Total hemoglobin}} = \frac{HbO_2}{HbO_2 + Hb}. \qquad (2)$$

Notably, tissue optics serves as the cornerstone of biophotonics and biomedical optics to deepen our knowledge of light-tissue interactions and develop noninvasive optical diagnostic methods and devices [89, 91].

To demonstrate the versatility of hyperspectral learning and hypercube recovery, we formulate two complementary machine-learning frameworks. In statistical learning (Fig. 2A), sampling of hyperspectral data in a subarea serves as a training dataset for hyperspectral learning. The hypercube of the entire area is reconstructed using a statistical learning algorithm (Methods). Fixed-design linear regression, featuring polynomial expansions, is utilized as the method of least-squares ($l_2$-norm minimization) [55, 62, 81, 83]. A full spectroscopic resolution is achieved in a range of Δλ = 0.5 – 1 nm, highly comparable to those of scientific spectrometers or spectrographs. This level of spectral resolution is 10 times better than the typical resolution of hyperspectral imaging of Δλ = 10 nm. Subsequently, the recovered hypercube is fitted using a tissue reflectance spectral model to extract the hemodynamic parameters (Methods) [85-90]. In deep learning (Fig. 2B), hyperspectral data in a subarea are directly fed into the tissue reflectance spectral model to compute the hemodynamic parameters within the same subarea. The obtained dataset serves to train a deep neural network that computes the hemodynamic parameters with RGB values (tristimulus) as an input. Specifically, the deep neural network is directly trained by using the hemodynamic parameters extracted from the tissue reflectance spectral model fed with the sampled data, thereby reducing the computational load. In both cases, separate training and validation datasets are employed to strengthen the learning algorithm for training: 80% of the data points among the sampled data (i.e., 600 data points out of 750) are randomly selected as a training dataset and the remaining 20% (i.e., 150 data points) are blindly tested as a testing dataset.

Importantly, deep learning informed by hyperspectral information is advantageous for designing explainable and interpretable neural networks. Among similar yet distinct terms, such as understandability and comprehensibility [77], spectrally informed deep learning enables transparency in the learning algorithm as it is understandable in a manner similar to statistical regression. The deep neural network model is designed to mimic the concept of polynomial expansions in statistical hyperspectral learning (Fig. 2C, Fig. S3, and Methods). In particular, the first hidden layer, which is one of the important hyperparameters for building a neural network, is fully connected with a relatively large number of nodes (18 nodes) to mimic hyperspectral learning, which transforms RGB values to a spectral intensity profile with a high spectral resolution. After the network is trained, each node in the first hidden layer possesses a distinct weight representing hyperspectral information, such that the RGB values of a certain hemodynamic parameter (e.g., $sPO_2$) generate the corresponding spectral feature, which is further propagated throughout the network (Fig. 2D). Figure 2E illustrates two representative cases of hyperspectral data measured from a tissue phantom by varying $sPO_2$ between $HbO_2$ and Hb (Fig. S3 and Supplementary Methods). The output values at different nodes in the first hidden layer can be understood based on the spectral intensity differences as a function of λ (Fig. 2E, F) that a scientific spectrometer or spectrograph can quantify. This direct spectral understanding should be differentiated from other conventional heatmaps or saliency maps employed for visualizing or explaining features extracted through typical convolutional neural networks [92, 93]. In particular, the differences in the computed output values of the first hidden layer between the two different RGB values of $HbO_2$ and Hb in a tissue phantom (Supplementary

Methods) resemble the spectral intensity differences between $HbO_2$ and $Hb$ measured from the same tissue phantom (Fig. 2E, F).

We comprehensively validate the hyperspectral learning-based instantaneous imaging method using an experimental model of vascular development. Chicken embryos serve as an excellent model system for imaging the development of blood vessel formation and hemodynamics (Fig. 3A) [94, 95]. The key statistical assumptions between the sampled area (i.e., the central line) and the entire image area are tested by evaluating the probability distribution profiles of each RGB channel (Fig. 3B) and the Q-Q plots (Fig. 3C). The sampled dataset (both hyperspectral and RGB data along the central line) is used to train a learning algorithm that has the input of RGB values and the output of hemodynamic parameters (i.e., $HbO_2$ and $Hb$). The residuals between the ground-truth (acquired by a pushbroom-type hyperspectral imaging system in Methods) and reconstructed hyperspectra are minimal, as shown with 95% confidence intervals as a function of $\lambda$ (Fig. 3D, E, G, and Methods). The hemodynamic parameter extractions are further supported by the 95% confidence intervals of residuals between the ground-truth and fitted hyperspectra obtained using the tissue reflectance spectral model (Fig. 3D, F, H, and Methods). Fig. 3G, H show that the reconstructed and fitted spectra are in excellent agreement with the ground-truth spectra. Moreover, the spectral angle mapper (Methods) between the ground-truth and reconstructed hypercubes captures a high degree of similarity with an average value of 0.035 rad (Fig. 3I). The direct comparison (i.e., spectral angle mapper) between the ground-truth and fitted hypercubes also return an average value of 0.046 rad (Fig. 3J), supporting the performance of hyperspectral learning and tissue reflectance spectral modeling.

Figure 4 depicts the hemodynamic maps of $Hb$, $HbO_2$, and $sPO_2$ corresponding to a white leghorn chicken (*Gallus gallus domesticus*, Hy-Line W-36) embryo on day 8 (Fig. 4A-C). In this statistical learning framework, after recovering the hypercube for the entire area via a hyperspectral learning algorithm, the recovered hypercube is fitted by using the tissue reflectance spectral model (Fig. 2A and Methods), thereby computing the hemodynamic parameters in a pixel-by-pixel manner. Figure. 4D clearly demonstrates the high levels of $sPO_2$ in the major blood vessel compared with capillary vessels. As a deep-learning framework (Fig. 2B), we design a fully connected deep neural network that takes RGB values as the input and returns the hemodynamic parameters as the output (Fig. 2C, Fig. S3, and Methods). As stated previously, the two unique design aspects are that the neural network encompasses the concept of polynomial expansions in hyperspectral learning and incorporates the output parameters extracted from the tissue reflectance spectral modeling (Fig. 2D and Methods). As illustrated in Figure 4E, the hemodynamic parameters are directly computed using the deep neural network. In particular, the hemodynamic maps obtained from both frameworks are in excellent agreement with the ground-truth hemodynamic maps generated via conventional pushbroom-type hyperspectral imaging, which are supported by high values of structural similarity index (Fig. 4F and Table S2). Interestingly, the conventional hemodynamic maps are noisier due to the motion artifact of the live sample, which results from the slow rate of data acquisition (data acquisition time = 45 minutes) and the mechanical scanning, as shown by the horizontal lines in Figure 4F.

Moreover, another advantage offered by learning-based snapshot imaging is that a video can be transformed into parameter maps after the learning algorithm is trained. Essentially, a video comprises a time series of multiple RGB images and each frame represents an RGB image. A video captured using a common trichromatic camera (e.g., a smartphone camera) can conveniently be processed using hyperspectral learning. Recently, smartphone models are capable of slow-motion video mode, featuring as high frame rates as 960 – 1920 frames per second (fps). Thus, hyperspectral learning enables unprecedentedly fast hyperspectral imaging with high temporal resolution (e.g. 0.0005 – 0.0010 sec). To demonstrate ultrafast dynamic imaging, we take advantage of an off-the-shelf smartphone (Samsung Galaxy S21) to record a video at 960 fps (temporal resolution = 0.0010 sec). Each frame representing an RGB image of the video is fed into the hyperspectral learning algorithm to compute the corresponding hemodynamic images. Then the hemodynamic images are restacked into a video format (Methods, and Supplementary Methods). Figure 5 shows time series hemodynamic maps of $sPO_2$ changes during a selected short period (Video S1, S2 for the entire duration of 0.5 sec) in a tissue phantom that has a rapid oxygen exchange process (Supplementary Methods and Fig. S3). The hemodynamic video with the millisecond resolution reveals rapid hemoglobin dispersion and oxygen diffusion, varying $sPO_2$ between $HbO_2$ and Hb. It should be noted that hyperspectral learning enables ultrafast dynamic imaging merely using a commercially available unmodified smartphone.

To further demonstrate dynamic imaging of peripheral microcirculation at a particular video frame rate, we record a video at 60 fps for a duration of 180 seconds (temporal resolution = 0.0167 sec) (Video S3 and Fig. S5, S6). As a model system for peripheral microcirculation in humans, we visualize spatiotemporal hemodynamic changes in the microvessels of the inner eyelid (i.e., the palpebral conjunctiva) (Fig. 6A, B). The inner eyelid is an easily accessible and highly vascularized peripheral tissue site that receives blood from the ophthalmic artery. Thus, the inner eyelid serves as a feasible sensing site for various diseases and disorders [96-98]. Figure 6C, D show the peripheral hemodynamic maps of Hb, $HbO_2$, and $sPO_2$ obtained during the resting state of a healthy adult volunteer. Videos of hemodynamic maps computed from the smartphone video are generated using the two complementary machine-learning frameworks of statistical learning and informed deep learning (Methods and Video S1-S3). In Figure 6, the $sPO_2$ maps reveal spatially complex patterns of perfusion in the inner eyelid, which are not evident in the photo (i.e., the RGB image). The statistical learning and deep learning-based hemodynamic maps of $HbO_2$ and Hb are in excellent agreement with each other, thereby indicating the reliable extraction of hemodynamic parameters.

Furthermore, we validate the computed hemodynamic maps obtained via the smartphone video (Video S3) against the functional imaging signals obtained using a functional near-infrared spectroscopy system (fNIRS) (Methods and Supplementary Methods). In this *in vivo* imaging setting, simultaneously acquiring the concurrent hypercube from the same sensing site (i.e., inner eyelid in Fig. 7A) is practically unachievable using a conventional hyperspectral imaging system. In this study, the peripheral hemodynamic signals obtained from the smartphone video are compared with the concurrent fNIRS signals averaged from all brain regions (Fig. 7D) and the fingertip (Fig. 7F) of the same subject measured using an fNIRS neuroimaging system (Methods and Supplementary Methods). We focus on the existence of the same low-frequency oscillations of $HbO_2$ and Hb associated with non-neuronal circulatory signals. In particular, low-frequency oscillations of $HbO_2$ and Hb in a frequency range of 0.01 – 0.1 Hz are commonly present in functional imaging signals (e.g., blood oxygen level-dependent functional magnetic resonance imaging and fNIRS) [99, 100]. Specifically, a hemodynamic phase of approximately 180º between $HbO_2$ and Hb changes ($\Delta HbO_2$ and $\Delta Hb$) during the resting state of healthy adults conveys autonomic physiological information, potentially serving as a biomarker of circulatory dysfunction.

We track hemodynamic signals in a relatively large vessel (Fig. S7) in terms of the changes in $HbO_2$ and Hb ($\Delta HbO_2$ and $\Delta Hb$), which are computed via statistical learning as well as deep learning. Figure 7B, c characterize $\Delta HbO_2$ and $\Delta Hb$ obtained after filtering out the heartbeat-related signals using a bandpass filter (0.01 – 0.1 Hz) (Methods). Moreover, vector analyses reveal that the phase differences between $\Delta HbO_2$ and $\Delta Hb$ over time under the resting state are 180.9º and 180.0º for statistical learning and deep learning, respectively (Fig. 7B, C). When the concurrent resting-state fNIRS signals from the cortical areas of the brain and another peripheral site (fingertip) from the same healthy subject are analyzed, the time-averaged phase differences between $\Delta HbO_2$ and $\Delta Hb$ also return 184.3º and 184.8º, respectively (Fig. 7E, G). Particularly, these consistent phase differences corresponding to the brain and peripheral sites (i.e., inner eyelid and fingertip) indicate that low-frequency hemodynamic oscillations are non-neuronal and circulate to the different parts of the body. More importantly, these results validate the accurate and reliable extraction of hemodynamic parameters from smartphone video recordings.

**Discussion**
The two machine learning frameworks of statistical learning and informed deep learning are complementary. In statistical learning, $l_2$-norm minimization (i.e., least squares regression) of hyperspectral learning enables stable and robust hypercube recovery with high spectral and temporal resolutions from an RGB image. This method is analogous to compressed sensing (i.e., $l_1$-norm minimization) such that incomplete measurements can be used to recover an entire image. The recovered hypercube is an intermediate product to be processed to extract key features or classifications of interest, requiring an additional step. On the other hand, in deep learning, the sampling of hyperspectral data in a subset area facilitates the direct incorporation of tissue optics modeling into the algorithm, resulting in direct parameter extractions. In addition, this spectrally informed deep-learning approach is advantageous for featuring a transparent network architecture, realizing explainable and interpretable learning. Moreover, this reported snapshot ability will offer ample opportunities for applying hyperspectral imaging to emerging areas where instantaneous data acquisitions have been the bottleneck.

Hyperspectral learning from an RGB image can potentially be enhanced by using the state-of-the-art three-color image sensors and high dynamic range images. First, conventional three-color sensors interpolate missing color information in selected pixel positions using a demosaicing algorithm, because each pixel records only one of the RGB colors in a format of a grid or checkerboard (also known as Bayer filter). Foveon X3 sensors capture all incident light (three colors) within the same pixel without having three separate color channels [101]. An image sensor with a color router can directly detect the entire color content without using a filter to select the intended color channel [102, 103]. Second, a higher color depth (also known as bit depth) in each pixel can further improve hyperspectral learning to avoid metamerism such that different spectral profiles result in the same RGB values. Because the color depth in each R, G, and B channel is 10-bit in our case, the number of different colors (combinations of RGB values) is $2^{10} \times 2^{10} \times 2^{10}$ (= 1.07 billion) colors. In other words, it is extremely unlikely that metamerism occurs. The higher the color depth of an RGB image, the more colors the image can store, further reducing the possibility of metamerism in hyperspectral learning.

In conclusion, the main advantages of this snapshot imaging method are the independence between the spectral and spatial resolutions, the hardware simplicity that could enable the use of commercially available smartphones, and the versatility offered by different machine learning techniques. We further envision that spectrally informed hyperspectral imaging will lay the foundation for mHealth technologies to noninvasively access hemodynamic information from an easily accessible peripheral tissue site using off-the-shelf smartphones in resource-limited settings or at-home settings.

**Materials and Methods**
*Dual-channel imaging setup mounted with a smartphone camera*
First, we constructed a dual-channel imaging spectrograph setup to acquire an RGB image in the entire field-of-view and hyperspectral data in a subarea; a line-scan hyperspectral system was mounted with a trichromatic camera (i.e., RGB camera and smartphone camera). This configuration was inspired by an image-guided hyperspectral line-scan system mounted on a telescope, which is a common configuration used in amateur astronomy [80]. In this dual-channel setup, a photometric slit, which is a mirror with a slit opening (slit width = 23 µm), enabled the simultaneous acquisition of an RGB image and hyperspectral line-scan data. An image reflected by the photometric slit (mirror portion) was captured using a trichromatic (RGB) camera (GS3-U3-120S6C-C, FLIR). The light passing through the slit was dispersed by a diffraction grating (groove density = 150 mm$^{-1}$) and captured using a monochrome camera (GS3-U3-120S6M-C, FLIR). A tungsten halogen light source (HL-2000-HP Light Source, Ocean Insight) was used as a broadband illumination source. Spectral calibration of the spectrograph was performed using a xenon calibration light source emitting multiple narrow peaks at specific wavelengths. An objective lens (4× and NA = 0.10) was mainly used to image the microvessels and the field-of-view was as small as 2.5 mm × 2 mm with a spatial resolution of 55 μm (Fig. S1). The spectral resolution was determined via a helium-neon laser that emits monochromatic radiation with a narrow bandwidth at 543.5 nm. The spectral range of hyperspectral line-scan data was 380 – 720 nm, and the spectral resolution Δλ was 0.5 nm (Fig. S2). The typical sizes of the RGB image and hyperspectral data were 750 × 300 pixels and 750 × 1 pixels in the spatial domain, respectively. Second, we incorporated a smartphone (Samsung Galaxy S21) into the dual-channel system using a cube beamsplitter (Supplementary Methods and Fig. S8). The color depths (bit depth) were 10-bit (file format: DNG) for photos and 8-bit (MP4) for videos, respectively. The smartphone's video-recording function (i.e., ultraslow-motion recording) was used to acquire videos at frame rates of 60 or 960 fps (Video S1-S3, Supplementary Methods, and Fig. S9). Third, we employed a motorized linear stage (T-LSR150A, Zaber) to move the entire system to directly acquire the ground-truth hyperspectral image cube (hypercube) in the entire field-of-view in the same manner as the pushbroom-type hyperspectral imaging system. A mechanical linear scan step was performed at 10 μm. Mechanical scanning was performed only for the validation studies. The data acquisition was synchronized using a custom-built MATLAB interface.

*Frontend data processing*
We implemented frontend data processing for both hyperspectral and RGB imaging to factor out the spectral contributions from the illumination, system responses, and background ambient light. The measured spectral intensity $I_M(\lambda)$ reflected from the sample is expressed as a function of wavelength λ:

$$I_M(\lambda) = L(\lambda)\, C(\lambda)\, S(\lambda)\, I_R(\lambda), \tag{3}$$

where $L(\lambda)$ denotes the spectral shape of the illumination light source, $C(\lambda)$ represents the spectral response of all the optical components in the system, $S(\lambda)$ depicts the spectral response function (also known as the sensitivity function) of the image sensors, and $I_R(\lambda)$ symbolizes the intensity reflected in the sample [26, 30, 53-65]. First, a white reflectance standard (SRS-99-010, Labsphere) was used to compensate for the spectral response of the system, including the illumination. The white reflectance standard possesses a reflectivity of 99% in the visible range, and its measured intensity can be expressed as:

$$I_{White}(\lambda) = L(\lambda)\, C(\lambda)\, S(\lambda). \tag{4}$$

Second, a black reflectance standard (SRS-02-010, Labsphere) with a reflectivity of 2% in the visible range was acquired to eliminate the ambient stray and background room light. After subtracting the black standard intensity $I_{Black}$ from each measurement, the normalized intensity reflected from the sample can be obtained:

$$I_R(\lambda) = \frac{I_M(\lambda) - I_{Black}(\lambda)}{I_{White}(\lambda) - I_{Black}(\lambda)}. \tag{5}$$

*Hyperspectral statistical learning*
We established a statistical learning framework to reliably reconstruct a hyperspectrum from RGB values. Conceptually, hyperspectral learning solves a pseudoinverse of $S_{3\times k}$ in equation (1) where $S_{3\times k}$ is the spectral response function in the R, G, and B channels of the camera. As the method of least-squares ($l_2$-norm minimization), we took advantage of fixed-design linear regression with polynomial expansions to solve $[S_{3\times k}]^+$ [55, 62, 81, 83]. First, the sampled hyperspectral and RGB data (*m* = training data size) were used to change this underdetermined problem to an overdetermined problem such that $X_{3\times m}$ and $Y_{k\times m}$ were formed by adding $x_{3\times 1}$ and $y_{k\times 1}$ of the training data. The relationship in equation (1) is re-written:

$$X_{3\times m} = S_{3\times k} Y_{k\times m}. \tag{6}$$

The relationship in equation (6) can be expressed in the following inverse format:

$$Y_{k \times m} = T_{k \times 3} X_{3 \times m}, \quad (7)$$

where $T_{k \times 3} = [S_{3 \times k}]^+$. Second, to incorporate the nonlinearity between RGB and hyperspectral data and improve the hyperspectral learning, we added 4th degree polynomial expansions by including a bias term such that $X_{3 \times m}$ was expanded to $\widehat{X}_{p \times m}$ with $p$ = 34 [55, 62, 83]:

$$\widehat{X}_{p \times m} = [1, 1, 1, R, G, B, R^2, G^2, B^2, RG, GB, RB, \ldots$$
$$R^3, G^3, B^3, RG^2, RB^2, GR^2, GB^2, BR^2, BG^2, RGB, \ldots$$
$$R^4, G^4, B^4, R^3G, R^3B, G^3R, G^3B, B^3R, B^3G, R^2G^2, R^2B^2, G^2B^2, R^2GB, G^2RB, B^2RG]^T. \quad (8)$$

Subsequently, equation (7) becomes:

$$Y_{k \times m} = \widehat{T}_{k \times p} \widehat{X}_{p \times m}. \quad (9)$$

Third, the inverse of the expanded matrix $\widehat{T}$ can easily be solved via QR decomposition or the Moore-Penrose pseudo-inverse [55, 62, 81-83]. Finally, upon substituting the RGB values ($x = [R, G, B]^T$) into equation (7), the corresponding hyperspectral intensity values ($y = [I(\lambda_1), I(\lambda_2), \ldots, I(\lambda_k)]^T$) were obtained. Thus, $\widehat{T}$ transforms RGB data to hyperspectral data.

*Tissue reflectance spectral modeling*
We make use of a tissue reflectance spectral model to extract key hemodynamic parameters from the ground-truth and recovered hyperspectral data. Light propagation in tissue can be modeled om accordance with the theory of radiative transport and robust approximations (e.g., diffusion, Born, and empirical modeling) [85-90]. Specifically, we conducted parameter extractions using an extensively used empirical modeling method. The intensity reflected from a biological sample can be expressed as a function of λ in the visible range:

$$I_R(\lambda) = \left[ b_1 \left( \frac{\lambda}{\lambda_0} \right)^{b_2} + b_3 \left( \frac{\lambda}{\lambda_0} \right)^{-4} \right] \exp\left[-b_4 \times \{b_5 \times \varepsilon_{HbO_2}(\lambda) + (1 - b_5) \times \varepsilon_{Hb}(\lambda)\}\right], \quad (10)$$

where $b_1$, $b_2$, and $b_3$ are associated with the scattering (Mie or Rayleigh) contributions at $\lambda_0$ = 800 nm, $\varepsilon_{HbO_2}(\lambda)$ denotes the absorption coefficient of oxygenated hemoglobin (HbO$_2$), $\varepsilon_{Hb}(\lambda)$ denotes the absorption coefficient of deoxygenated hemoglobin (Hb), $b_4$ is the hemoglobin concentration multiplied by the optical pathlength, and $b_5$ is the blood oxygen saturation (sPO$_2$). In the chicken embryo study, the lipid content of the egg yolk is high, as avian embryos rely on over 90% of the caloric requirement of fatty acids [104]. Thus, an additional absorption coefficient term for lipid was added to equation (10) to account for this high lipid content [105]. All fitting parameters were computed using the simplex search (Nelder-Mead) algorithm [106]. The differences (residuals) between the original and fitted hyperspectra were plotted with 95% confidence intervals as a function of λ (Fig. 3, and Fig. S5, S6).

*Deep neural network designing and training*
We designed a deep neural network considering the polynomial expansions of hyperspectral learning (equation (8)) and tissue reflectance spectral modeling (equation (10)). First, as an alternative to the sequence of hyperspectral learning and hemodynamic parameter extraction in statistical learning, a deep neural network was established that utilized RGB values as the input and returned the hemodynamic parameters of HbO$_2$ and Hb (or total hemoglobin and sPO$_2$) as the output. Hemodynamic parameters were calculated using the tissue reflectance spectral model (equation (10)). The two hemodynamic parameters as outputs were fed to the network for supervised learning, resulting in rapid convergence and high prediction accuracy. The neural network was trained and validated using 80% (600 data points) and 20% (150 data points) of locally sampled data.

Second, in the fully connected network, 18 nodes (neurons) in the first hidden layer are intended to have optimal selections of RGB values with different weights on the R, G, and B values and their combinations. It should be noted that the first hidden layer is conceptually understandable in a manner similar to the polynomial expansions in statistical hyperspectral learning (equation (8)), thereby offering a transparent neural network model:

$$\begin{aligned} N_{1,1} &= w_{1,\text{R}}R + w_{1,\text{G}}G + w_{1,\text{B}}B, \\ N_{1,2} &= w_{2,\text{R}}R + w_{2,\text{G}}G + w_{2,\text{B}}B, \\ &\vdots \\ N_{1,i} &= w_{i,\text{R}}R + w_{i,\text{G}}G + w_{i,\text{B}}B, \end{aligned} \quad (11)$$

where $R$, $G$, and $B$ are the input RGB values, $N_{1,i}$ indicates the $i^{\text{th}}$ node of the first hidden layer, and $w_{i,\text{RGB}}$ is a weight connecting the R, G, or B input nodes and the $i^{\text{th}}$ node of the first hidden layer, respectively. Overall, the network included the input layer with three nodes, four hidden layers, and the output layer (Fig. S3). Batch normalization, the softplus activation function (i.e., a smooth approximation to ReLU), and the ADAM optimization were employed for the network. The weight decay (factor for $l_2$ regularization) was specified to be $10^{-5}$ to avoid overfitting. The initial learning rate was set to 0.01, but the learning rate was set to drop for a given number of epochs. The batch size was 20 and the value of the epoch was 15 (Fig. S3).

*Spectral angle mapper*
The spectral angle mapper directly compares two spectra (e.g., measured spectrum and reference spectrum) allowing for a geometric interpretation in a space with dimensionality equal to the number of discretized wavelengths (*k*). Specifically, the spectral angle mapper calculates an angle between the spectra in the *k*-dimensional space, quantifying spectral similarity in a pixel-by-pixel manner [107].

$$\alpha = \cos^{-1}\left( \frac{\sum_{i=1}^{k} I_{\text{A}}(\lambda_i) I_{\text{B}}(\lambda_i)}{\sqrt{\sum_{i=1}^{k} I_{\text{A}}(\lambda_i)^2} \sqrt{\sum_{i=1}^{k} I_{\text{B}}(\lambda_i)^2}} \right), \quad (12)$$

where $I_\text{A}$ is the spectrum to be compared and $I_\text{B}$ is the reference spectrum.

*Functional near-infrared spectroscopy (fNIRS) measurements*
We utilized an fNIRS neuroimaging system (NIRx Medical Technologies) to concurrently measure hemodynamic signals from the brain and fingertips during the resting state of healthy adult volunteers. In brief, the head cap and finger clip had two sets of laser sources and detectors at 785 nm and 830 nm, respectively, and the separation distance between the laser and the detector was 30 mm. Six detection channels were used to cover all brain areas and two detection channels were employed for a fingertip. The sampling rate of fNIRS was 10 Hz. $HbO_2$ and Hb signals were synchronized with the dual-channel imaging setup and smartphone camera. The fNIRS signals were processed for motion correction and spatial smoothing with a full width at half maximum of a 5-mm isotropic Gaussian kernel. A bandpass filter (0.01 – 0.1 Hz, 3$^{\text{rd}}$ order Butterworth) was used to extract the low-frequency signals of ΔHbO$_2$ and ΔHb. The phase differences between the HbO$_2$ and Hb signals (ΔHbO$_2$ relative to ΔHb) were calculated after application of the bandpass filter [100].

**Funding**
This work was supported by the Technology Accelerator Challenge Prize from National Institute of Biomedical Imaging and Bioengineering, the John E. Fogarty International Center at the National Institutes of Health (R21TW012486), the Ralph W. and Grace M. Showalter Trust, Veterans Affairs Merit Award (I01BX005816), and Ministry of Trade, Industry, and Energy (MOTIE) in Korea (P0017306).

**Acknowledgments**
We thank Norvin Bruns for helping with the imaging setup and Darrin Karcher for providing the fertilized eggs (*Gallus gallus domesticus*, Hy-Line W-36) at the Purdue University Animal Sciences Research and Education Center poultry unit.

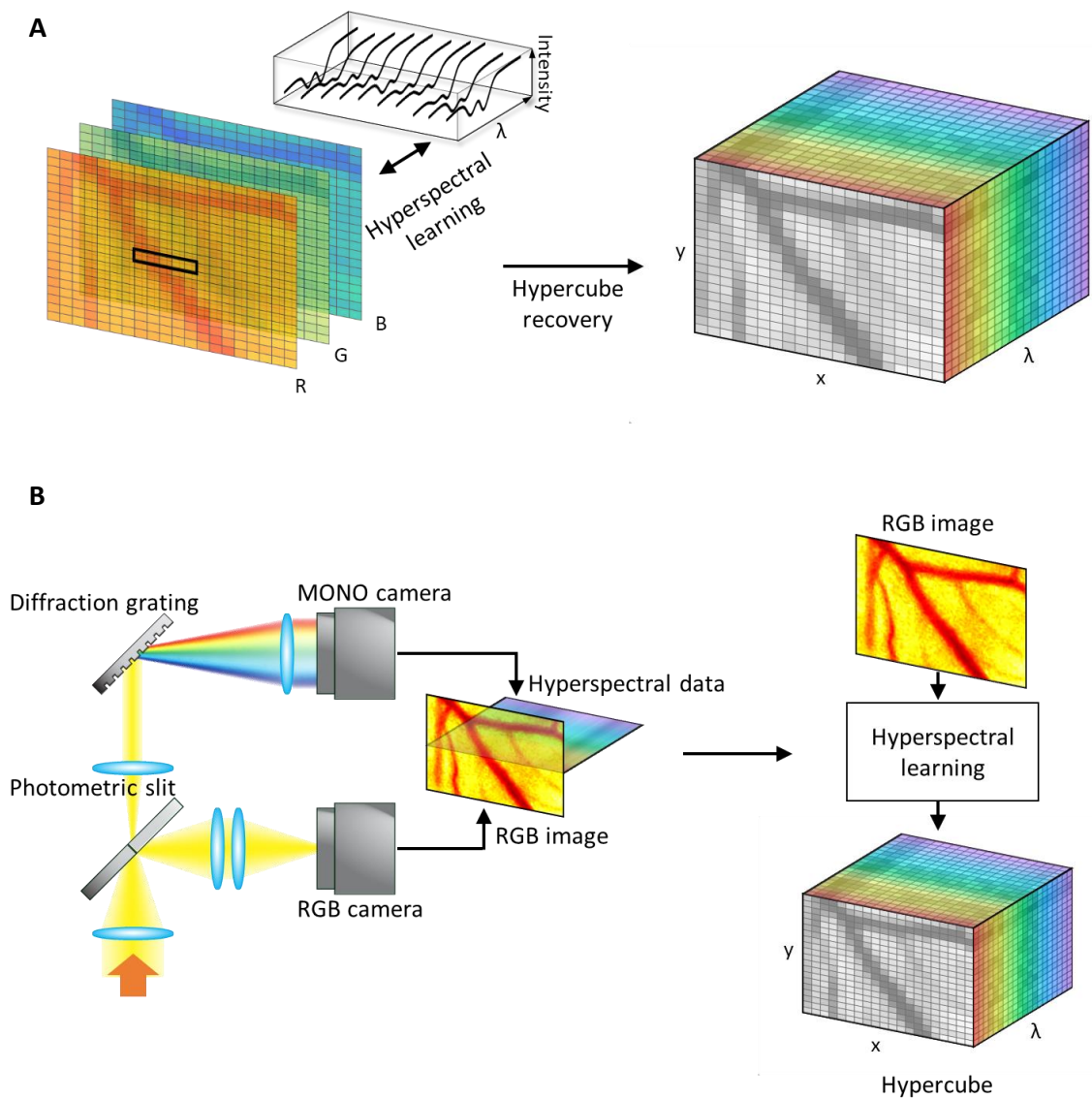

**Figure 1. Schematic illustration of hyperspectral learning for instantaneous imaging using a conventional trichromatic camera (i.e., three-color image sensor).** A sampling of detailed spectral (hyperspectral) information in a small yet representative subarea enables hyperspectral learning to recover a hypercube in the entire image as well as to perform spectrally driven machine learning for physical or biological parameter extractions. **(A)** Hyperspectral learning for hypercube recovery. Hyperspectral data in a small subarea are used to train a learning algorithm that takes the RGB values in each pixel as the input and returns a spectrum with a high spectral resolution (also known as spectral super-resolution). Hyperspectral learning is not affected by an intrinsic tradeoff between spatial and spectral resolutions, which often limits conventional snapshot hyperspectral imaging. **(B)** Subarea sampling of hyperspectral data. A dual-channel spectrograph with a photometric slit (Methods) simultaneously acquires an RGB image in the entire field-of-view and a hyperspectral line scan in a small subarea (i.e., the central line) in a single-shot manner. The sampled hyperspectral data serve as prior information or physical constraints for training the hyperspectral learning algorithm. The spectrally driven informed machine learning algorithm trained by the sampled hyperspectral data transforms the RGB image

into a hyperspectral image dataset (also known as a hypercube), which can further be used to extract physical or biological parameters.

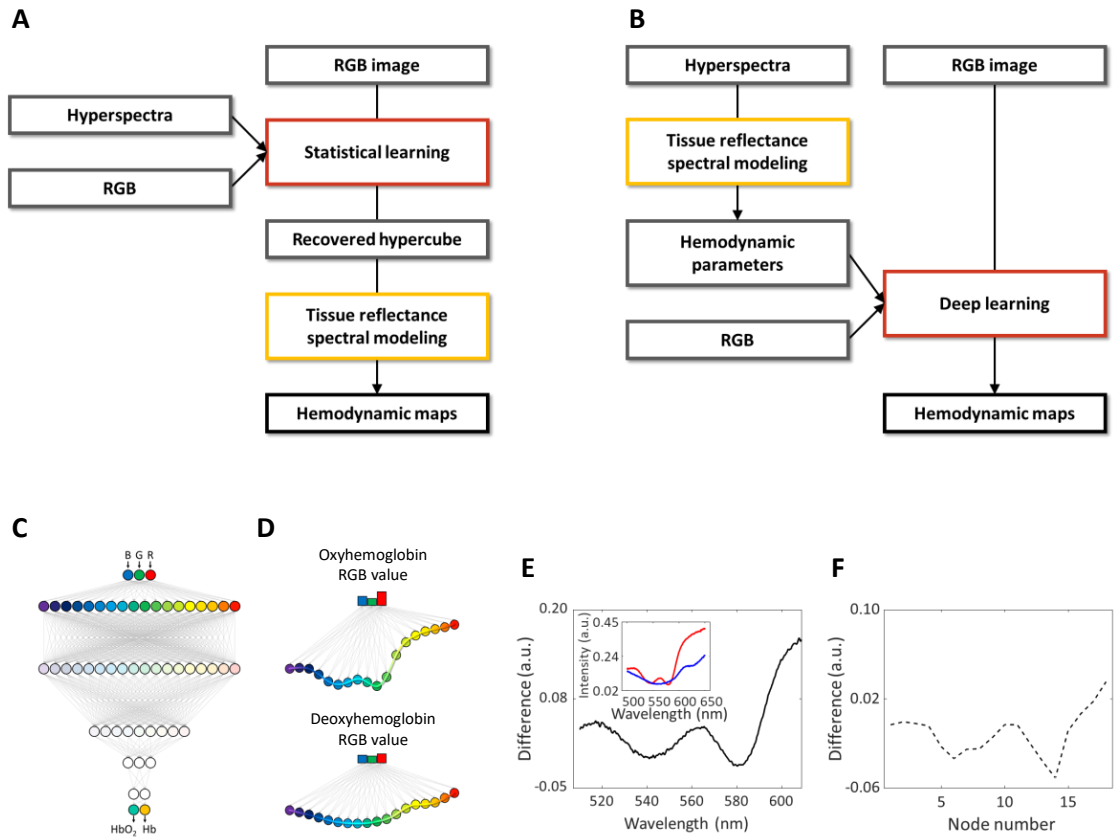

**Figure 2. Flowcharts of two complementary machine learning frameworks for hypercube recovery and hemodynamic imaging and schematic illustration of deep learning networks.**
**(A)** Statistical learning involving hyperspectral learning and tissue reflectance spectral modeling. Hyperspectral and RGB data in a subarea are used to train a statistical learning algorithm. The trained statistical learning algorithm is subsequently applied to the RGB data in the entire area to reconstruct the hypercube. Using tissue reflectance spectral modeling, the key hemodynamic parameters are extracted from the hypercube to generate the hemodynamic maps. **(B)** Deep learning informed by the tissue reflectance spectral model. The hemodynamic parameter values (output) from the hyperspectral data in the subarea extracted via tissue reflectance spectral modeling of the corresponding RGB values (input) are used to train the deep-learning network. The trained deep-learning network is thereafter applied to the RGB data for the entire area to generate the hemodynamic maps. **(C)** Illustration of the deep neural network that takes RGB values and returns key hemodynamic parameters (i.e., $HbO_2$ and Hb). This network is trained by the RGB values and hemodynamic parameters in a subarea; $HbO_2$ and Hb values are computed from the ground-truth hyperspectral data obtained with conventional pushbroom-type hyperspectral imaging system in the subarea via the tissue reflectance spectral model (Methods). **(D)** The first hidden layer is fully connected to 18 nodes (or neurons) to mimic the polynomial expansions of statistical learning. **(E)** Representative spectral intensity differences in the ground-truth hyperspectral data obtained from a tissue phantom with oxygenated and deoxygenated hemoglobin. Inset: The ground-truth hyperspectral data when the sample is oxygenated (red curve) and deoxygenated (blue curve) (i.e., $HbO_2$ or Hb, respectively). **(F)** Representative difference in the computed output values of the first hidden layer between two different RGB values of $HbO_2$ and Hb from the same tissue phantom in **E**. The order of the nodes is assigned such that the rank of the output differences is the same as that of the wavelengths in the spectral intensity differences.

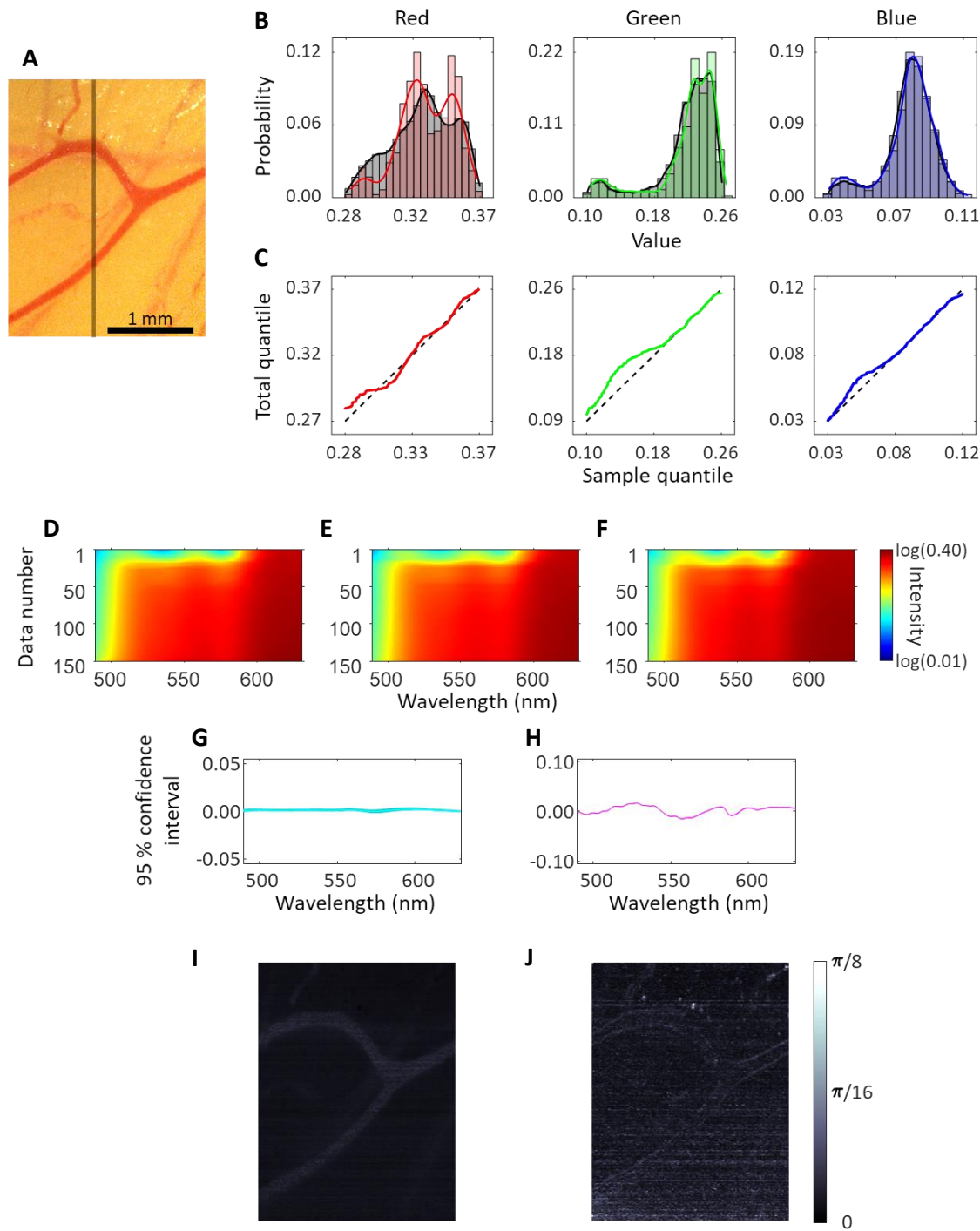

**Figure 3. Comprehensive evaluations of hyperspectral learning, hypercube recovery, and hemodynamic parameter extractions in an experimental vascular developmental model.**
**(A)** Photograph (RGB image) of the chicken embryo on day 8. The vertical gray line represents a subarea for hyperspectral sampling. **(B)** Probability distributions and histograms of R, G, and B values of the entire image area and the subarea. The black bins show the distributions of R, G, and B values of the entire image area. The black solid lines are displayed together with the black bins to visualize the trends of the data distributions of the entire image area. The colored bins depict the distributions of R, G, and B values of the subarea (i.e., the gray line in **A**) according to the RGB colors. The colored lines are illustrated together with the colored bins to visualize the trends in the data distribution of the subarea. **(C)** Quantile-quantile (Q-Q) plots for distribution comparisons between the entire field-of-view and the subarea. The subarea (gray line in **A**) ensures a similar intensity distribution as the intensity distributions of the entire area (**A**) in each RGB channel. The black dotted straight diagonal line indicates that the two distributions are identical. **(D, F)** Visualization of the ground-truth (**D**), reconstructed (**E**), and fitted (**F**) spectra from the testing data (data size = 150) of the sampled data (data size = 750). A logarithmic scale of intensity is used for enhanced visualization. **(G)** Comparison between the ground-truth and reconstructed spectra in the testing dataset. The differences (residuals) between the ground-truth and reconstructed spectra are plotted with 95% confidence intervals for each $\lambda$. **(H)** Comparison between the ground-truth and fitted spectra in the testing dataset. Hemodynamic parameters are extracted using the tissue reflectance spectral model (Methods). The differences (residuals) between the ground-truth and fitted spectra are plotted with 95% confidence intervals. **(I)** Spectral angle mapping image between the ground-truth and reconstructed hypercubes. **(J)** Spectral angle mapping image between the ground-truth and fitted hypercubes.

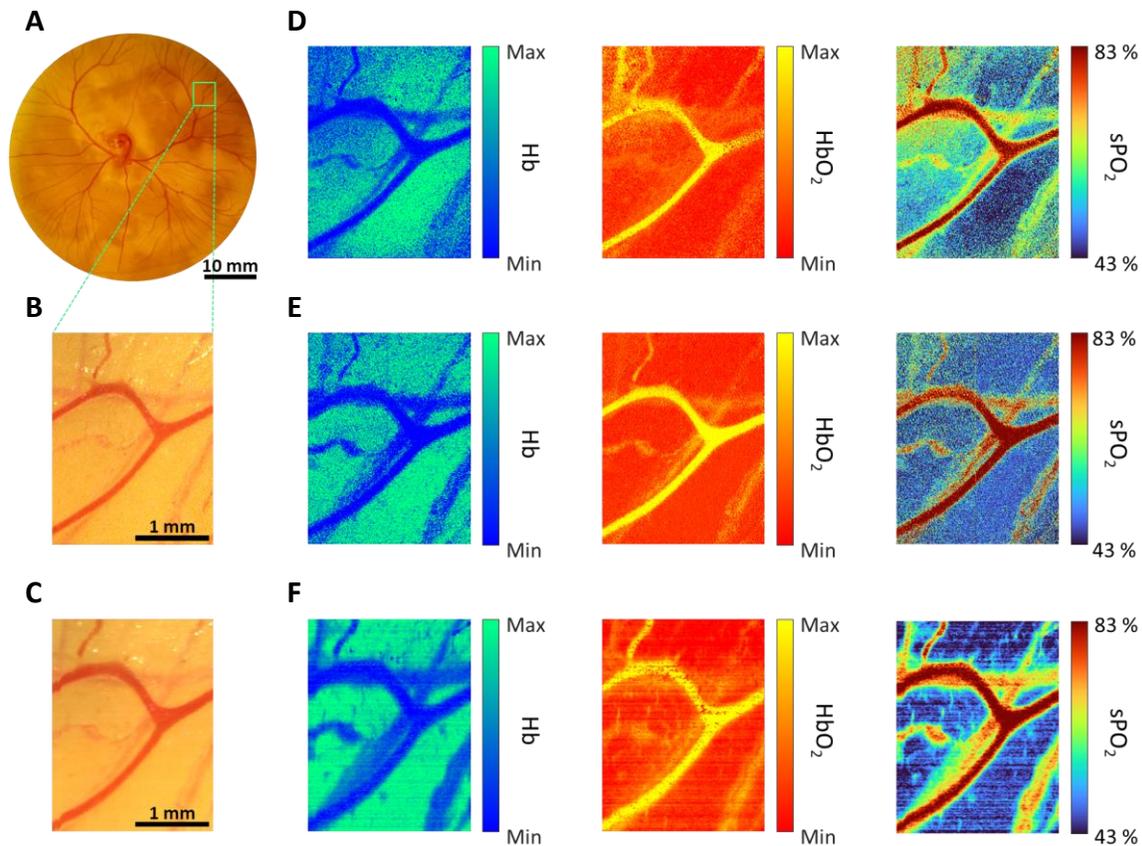

**Figure 4. Hemodynamic maps of oxygenated hemoglobin (HbO$_2$), deoxygenated hemoglobin (Hb), and oxygen saturation (sPO$_2$) in the experimental model of vascular development.** **(A)** Chick embryo in a petri dish as an experimental vascular developmental model. The image area for learning-based snapshot imaging with a high spatial resolution is marked by a green box. **(B)** RGB image in field-of-view (green box in **A**) for hyperspectral learning-based snapshot imaging. **(C)** RGB image generated by the conventional pushbroom-type hyperspectral imaging system. **(D)** Statistically learning-based hemodynamic maps. Hyperspectral learning generates a hypercube that is processed to compute hemodynamic maps using tissue reflectance spectral modeling. **(E)** Deep learning-based hemodynamic maps. Hemodynamic maps are directly generated using a deep neural network that takes the RGB values as the input and returns HbO$_2$ and Hb values as the output. **(F)** Reference hemodynamic maps generated by a conventional pushbroom-type hyperspectral imaging system for validation. The two learning-based hemodynamic maps (**D** and **E**) are in excellent agreement with the ground-truth maps (**F**). Statistical learning and deep learning-based hemodynamic maps are further assessed using the structural similarity index (Supplementary Methods) to show that they are qualitatively identical to the reference hemodynamic maps (Table S2).

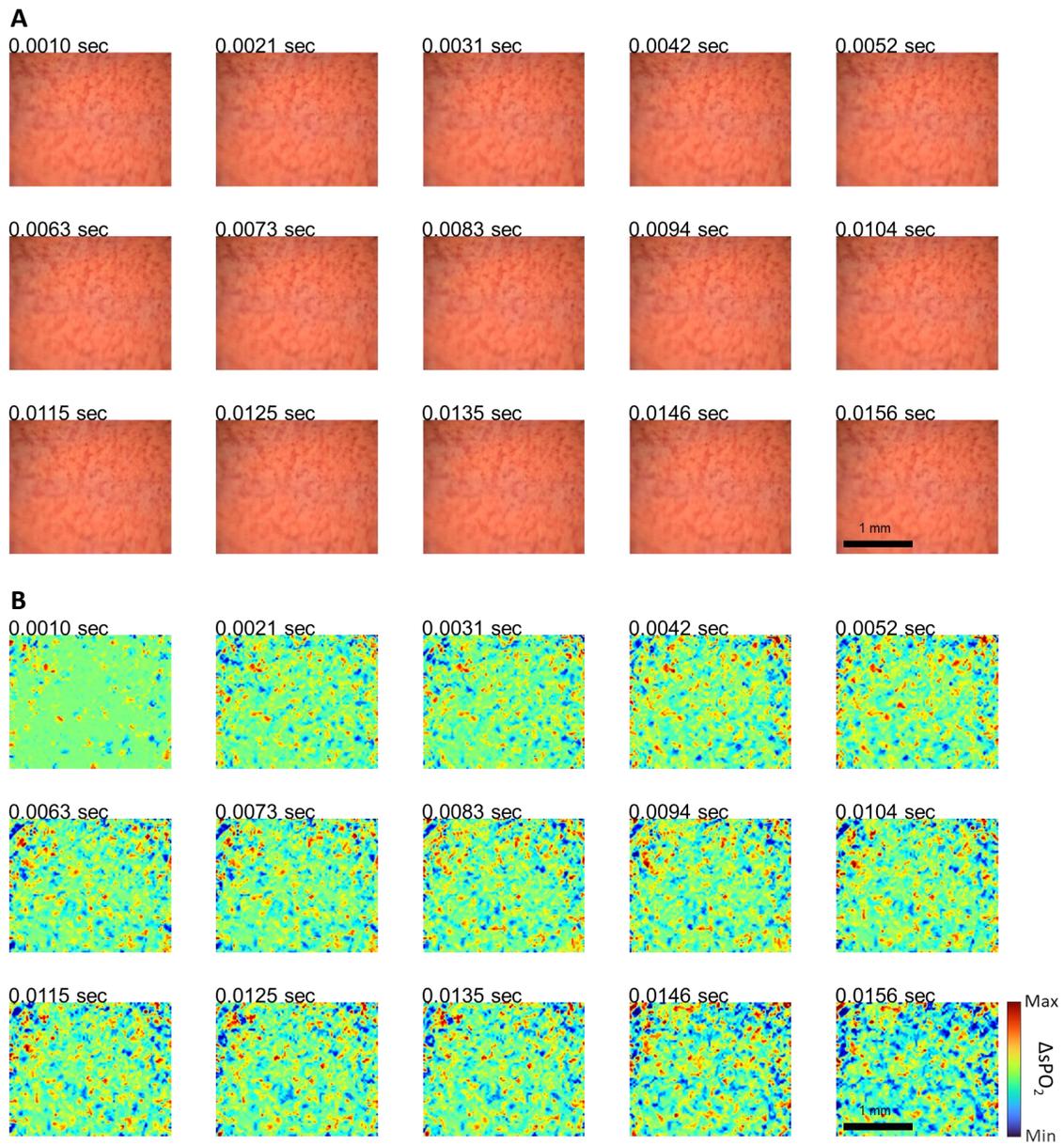

**Figure 5. Ultrafast dynamic imaging of sPO$_2$ changes via hyperspectral learning of smartphone's ultraslow video recording. (A)** Selected 15 frames (RGB images) from a video of the dynamic tissue phantom recorded by the smartphone's built-in ultraslow-motion mode (960 fps). **(B)** Hemodynamic change images of sPO$_2$ (ΔsPO$_2$) compared with the initial image, processed using the corresponding RGB images with the deep-learning algorithm. Hyperspectral learning allows an ultrafast data acquisition rate at a temporal resolution of 0.0010 sec (Video S1 for the entire duration) using the off-the-shelf smartphone.

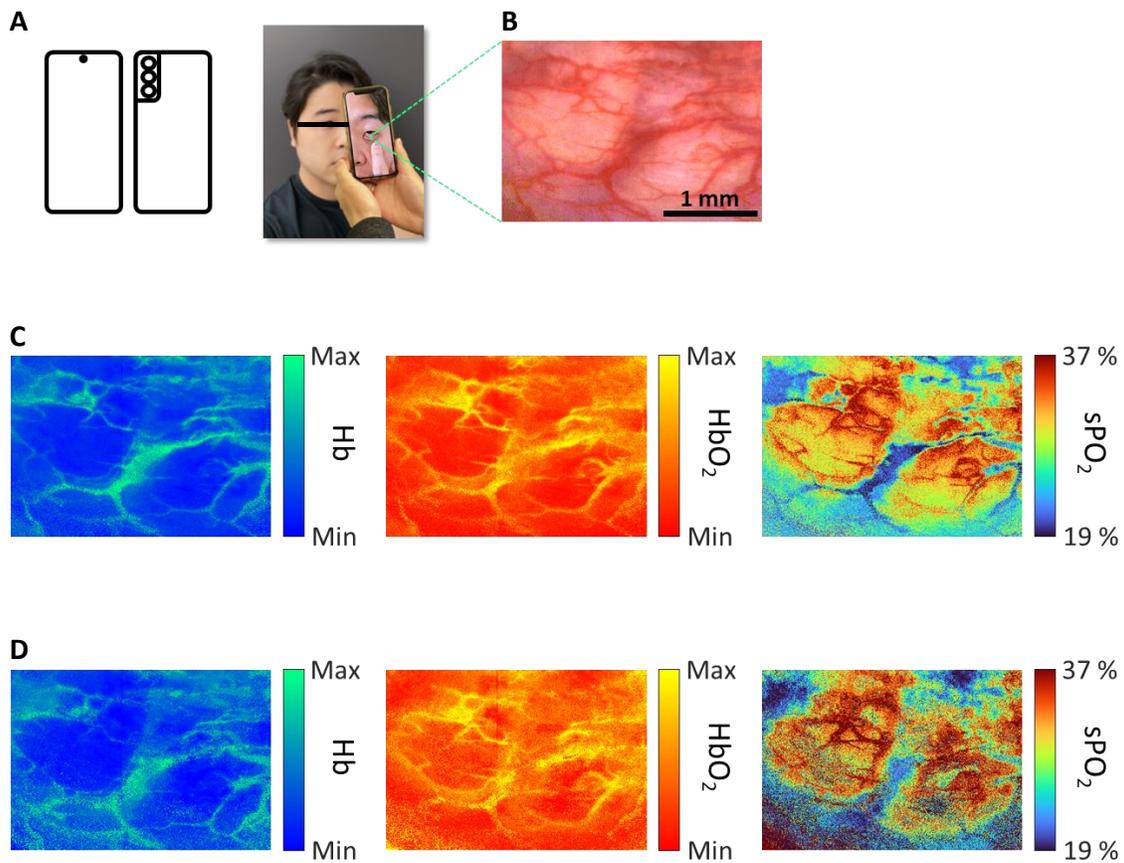

**Figure 6. Representative peripheral hemodynamic maps of HbO$_2$, Hb, and sPO$_2$ extracted from a healthy adult in the resting state.** The inner eyelid (i.e., palpebral conjunctiva) has high vasculature connected to the ophthalmic artery, serving as an easily accessible peripheral tissue site. **(A)** Photograph of a healthy adult volunteer taking a picture with a smartphone while the inner eyelid pooled down. **(B)** RGB image of the inner eyelid with high spatial resolution showing the field-of-view for hyperspectral learning-based snapshot imaging. Microvessels in the inner eyelid are clearly visible without the effects of skin pigments, which are easily accessible for imaging. **(C)** Statistical learning-based snapshot hemodynamic maps of the inner eyelid. Hemodynamic maps are generated using tissue reflectance spectral modeling based on spectral information from the recovered hypercube. **(D)** Deep learning-based snapshot hemodynamic maps of the inner eyelids. Hemodynamic maps are directly generated using the deep neural network, which returns HbO$_2$ and Hb values as the output when the RGB values are employed as the input.

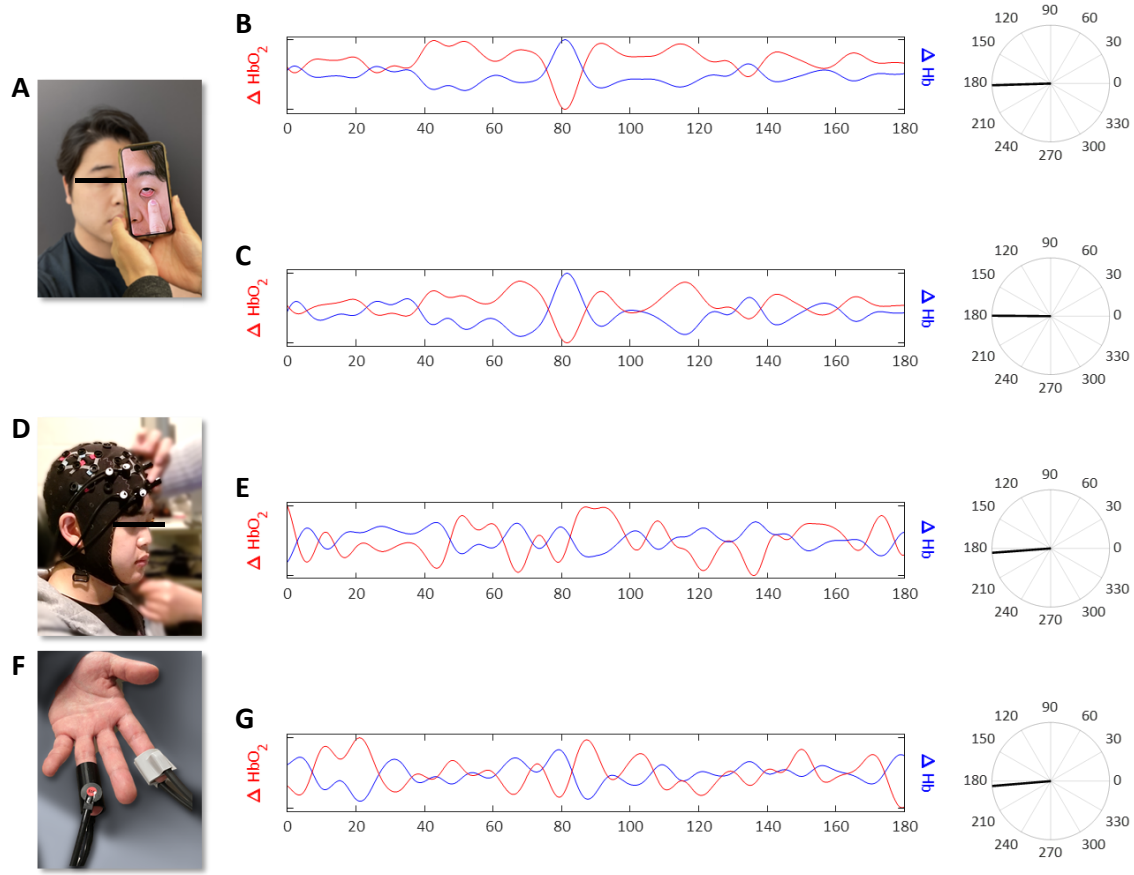

**Figure 7. Temporal changes in the eyelid's hemodynamic signals extracted from the smartphone video capture and concurrent resting-state functional near-infrared spectroscopy (fNIRS) signals from the brain and another peripheral site (fingertip). (A)** Representative photographs of data acquired from the inner eyelid of a healthy participant. A smartphone camera can easily record a video of the inner eyelid. (**B** and **C**), Temporal traces of statistical learning-based (**B**) and deep learning-based (**C**) hemodynamic signals of changes in $HbO_2$ ($\Delta HbO_2$) and Hb ($\Delta Hb$) in the inner eyelid. A relatively large blood vessel is selected (Fig. S7) to analyze the hemodynamic parameters. A temporal resolution of 0.016 sec is set by the smartphone video recording setting (Video S3). A vector representation of the time-averaged phase difference between the corresponding $\Delta HbO_2$ and $\Delta Hb$ signals in the eyelid is displayed on the right-hand side of the temporal trace. **(D)** Representative photographs of the data acquisition from the brain of the subject using fNIRS. The fNIRS signals from the brain of the same subject are also measured at the same time to be compared with the signal from the smartphone video. **(E)** Temporal traces of concurrent fNIRS signals averaged from all brain regions of the same subject with a vector representation of the time-averaged phase difference between $\Delta HbO_2$ and $\Delta Hb$ from all brain regions. **(F)** Representative photographs of the data acquisition from the fingertip of the subject using fNIRS. The fNIRS signals from the fingertip of the same subject are also measured at the same time to be compared with the signal from the smartphone video. **(G)** Temporal traces of concurrent fNIRS signals averaged from the fingertip of the same subject with a vector representation of the time-averaged phase difference between the corresponding $\Delta HbO_2$ and $\Delta Hb$ in the fingertip. The phase difference extracted from the smartphone video (**B** and **C**) is in excellent agreement with the functional imaging modality (**E** and **G**).

**Supplementary Methods**

*Tissue phantom preparation*
To demonstrate ultrafast hemodynamic imaging at a temporal resolution of 0.001 sec (960 frames per second), we prepared a tissue-mimicking phantom where oxygen depletion occurs in a rapid manner. The tissue phantom was composed of an aqueous suspension of polystyrene microspheres and lyophilized hemoglobin to mimic the light scattering and absorption properties of biological tissue, respectively. Specifically, the aqueous suspension contained 40% (v/v) 0.503 µm microspheres (PS03001, Bangs Laboratories) and 20% (w/v) lyophilized hemoglobin (H0267, Sigma-Aldrich) with a pH value of 7.4 at room temperature (23 ± 2 °C and 30 – 40% relative humidity). The scattering coefficient was 20 cm$^{-1}$ at 550 nm and the anisotropy factor was 0.88 at 550 nm, which were calculated using Mie theory [1, 2]. The tissue phantom was fully oxygenated initially because hemoglobin is about 100% saturated when the partial pressure of oxygen is 100 mmHg and the partial pressure of oxygen in the air is 160 mmHg at sea level [3]. To deplete oxygen of the tissue phantom in a certain area only, we randomly broadcasted a powder of sodium dithionite ($Na_2S_2O_4$ and also known as sodium hydrosulfite, Sigma-Aldrich) that deoxygenates hemoglobin without changing the native structure of hemoglobin (Fig. S3) [4]. To make a completely deoxygenated tissue phantom, 0.1 mg of sodium dithionite was added per 1 ml of the tissue phantom and was mixed gently not to make foams or bubbles. The random dispersion of sodium dithionite was imaged using the ultraslow video recording mode in Samsung Galaxy S21 (Video S1, S2).

*Sample preparation – Chick embryo*
As an experimental model of vascular development, we utilized chicken embryos (*Gallus gallus domesticus*, Hy-Line W-36). The temperature inside the incubators was set to 37 °C and the humidity was 55% supplied by filtered tap water. Before imaging, the eggs were placed on their side for one hour so that the embryo was raised. The shell was sterilized by wiping it with 70% ethanol. After the shell dried, the eggs were gently cracked and transferred the content to a petri dish for imaging.

*Structural similarity index calculation*
To validate the learning-based hemodynamic maps, we computed a structural similarity index to quantitatively compare with the hemodynamic maps generated using the conventional hyperspectral imaging system. Specifically, a structural similarity index between two images captures three characteristics of image luminance ($J_L$), contrast ($J_C$), and structural information ($J_S$) [5]. Three characteristics can be expressed:

$$J_L(P,Q) = \frac{2\mu_P\mu_Q + V_1}{\mu_P^2 + \mu_Q^2 + V_1}, J_C(P,Q) = \frac{2\sigma_P\sigma_Q + V_2}{\sigma_P^2 + \sigma_Q^2 + V_2}, \text{ and } J_S(P,Q) = \frac{\sigma_{PQ} + V_3}{\sigma_P\sigma_Q + V_3}, \tag{S1}$$

where $P$ and $Q$ denote two different hemodynamic maps, $\mu_P$ and $\mu_Q$ are the mean intensity over the entire image, $\sigma_P$ and $\sigma_Q$ are the standard deviation, and $\sigma_{PQ}$ is the covariance. The constant values $V_1$, $V_2$, and $V_3$ are included to avoid instability when the denominators are close to zero:

$$V_1 = (O_1 H)^2, V_2 = (O_2 H)^2, \text{ and } V_3 = V_2/2, \tag{S2}$$

where $O_1$ and $O_2$ are small constant values ($\ll 1$) and $H$ is the dynamic range of pixel values (e.g., 255 for 8-bit). By combining the three comparisons in equation (S1), we can calculate a structural similarity index between $P$ and $Q$:

$$\text{Structural similarity index }(P,Q) = J_L(P,Q) \cdot J_C(P,Q) \cdot J_S(P,Q). \tag{S3}$$

*Vessel and tissue area separations*

To isolate major blood vessel areas, we utilized a color image segmentation method to differentiate vessel and tissue areas in eyelid images and videos. Unlike the intensity in grayscale images, the RGB color values can make a complex segmentation problem simple [6]. We selected a set of RGB threshold values empirically to isolate blood vessel areas in the eyelid images (Fig.S7). The thresholds of RGB values were selected to create a binary segmentation mask as follows:

|  | R channel (min – max) | G channel (min – max) | B channel (min – max) |
|---|---|---|---|
| Blood vessel area | 0.29 – 1.00 | 0.14 – 1.00 | 0.14 – 1.00 |
| Avascular tissue area | 0.00 – 0.29 | 0.00 – 0.11 | 0.00 – 0.13 |

*Color correction of smartphone image*

Digital photographs acquired by different camera models exhibit distinguishable colors even though the same scenes are captured at the same time. This dissimilarity originates from the fact that the spectral response function (also known as the sensitivity function) is device dependent and significantly varies from model to model [7-9]. The color differences can also be related to the use of different color spaces and data formats [10]. Conventionally, white balancing is used to adjust colors, but it is not sufficient to correct color distortions. To ensure color consistency among the different cameras, we utilized a color correction method [7, 11-13]. The relationship of color values between two photographs can be expressed:

$$X_2 = MX_1, \qquad (S4)$$

where $M$ is a $3 \times 3$ conversion matrix and $X_1$ and $X_2$ are the reshaped RGB data captured using the two different imaging systems (i.e., smartphone camera and dual-channel imaging setup, respectively). Given a large number of pixels in our photographs, $M$ can easily be solved by the least squares regression of fixed-design (e.g., Moor-Penrose pseudo-inverse) as an overdetermined problem. Subsequently, the conversion matrix $M$ converted RGB values of a smartphone camera to RGB values of the dual-channel imaging setup (Fig.S8).

*Video Stabilization*

To minimize motion artifacts caused by the unintended movement of participants and the optical distortions of the lens system, we implemented a digital image stabilization method. Essentially, a video comprises a time sequence of images (i.e., frames). Thus, the registration and alignment of the whole frames in a video are necessary to remove motion artifacts or optical distortions [14]. The video stabilization method we implemented involving two steps; finding/tracking salient features in each frame and geometric conversion among frames. First, we utilized the speeded-up robust features (SURF) algorithm to detect salient features [15]. SURF is designed to identify salient features of objects, invariant to image rotation and scale with high computation speed and repeatability. SURF identified features related to blood vessels in the inner eyelid video (Fig.S9A-C). Second, we used the M-estimator sample consensus (MSAC) algorithm to formulate a geometric conversion between the different frames. The feature points in each frame served as anchors to register and align the different frames by adjusting the scale, rotation, and distortion (Fig.S9D) [16].

**Table S1. Comparisons of the recently reported snapshot hyperspectral imaging methods and systems**

| Methods | Spectral range (nm) | Spectral datapoints | Estimated spectral interval (nm)* | Number of pixels for imaging | Temporal resolution (ms) | Portability | Smartphone camera | Hemodynamic measurements | Reference |
|---|---|---|---|---|---|---|---|---|---|
| Computed tomographic imaging spectrometer | 450 – 700 | 50 | 5 | 208 × 208 | 3 | ○ | × | ○ | [17] |
| Two LEDs (multispectral imaging) | 470 and 530 | 2 | Not available | 128 × 128 | 9 | × | × | ○ | [18] |
| Filter wheel | 400 – 800 | 80 | 5 | 640 × 480 | 5,000 – 16,000 | × | × | ○ | [19] |
| Multispectral filter array | 481 – 632 | 16 | 9 | 256 × 512 | 50 | × | × | ○ | [20] |
| Spectral filters | 400 – 700 | 31 | 10 | 768 × 768 | 250 | × | × | × | [21] |
| Geometric phase microlens | 450 – 700 | 28 | 9 | 350 × 350 | Camera acquisition time | ○ | × | × | [22] |
| Spectral filter array | 386 – 898 | 64 | 8 | 28 × 20 | 1 – 13 | ○ | × | × | [23] |
| Diffraction grating filter | 400 – 700 | 102 | 3 | 384 × 384 | 18,000 | ○ | × | × | [24] |
| Metasurface tuned filter with metalens | 795 – 980 | 20 | 9 | Not available | Camera acquisition time | ○ | × | × | [25] |
| Filter array | 470 – 620 | 16 | 9 | 272 × 512 | 16 | ○ | ○ | ○ | [26] |
| Random spectral filters | 400 – 700 | 16 | 19 | 480 × 640 | > 8,000 | ○ | ○ | × | [27] |
| Spectrally modulated polarimetry | 450 – 750 | 56 | 3 | 400 × 400 | Camera acquisition time | ○ | × | × | [28] |
| Random filter array of Fabry–Pérot | 450 – 650 | 64 | 10 | 640 × 480 | 31 | ○ | × | × | [29] |
| Hyperspectral learning | 485 – 650 | 140 | 1 | 5960 × 5100 | 1 (960 fps) and camera acquisition time | ○ | ○ | ○ | This paper |

*The spectral interval is calculated by a ratio of the spectral range to the spectral data points because the information on the exact spectral resolution is often omitted.

**Table S2. Structural similarity index values to compare learning-based hemodynamic maps with a conventional pushbroom-type hyperspectral imaging system**

|  |  | Hyperspectral learning - Deep learning | | | Hyperspectral learning - Statistical learning | | |
|---|---|---|---|---|---|---|---|
|  |  | Oxygenated hemoglobin | Deoxygenated hemoglobin | Oxygen saturation | Oxygenated hemoglobin | Deoxygenated hemoglobin | Oxygen saturation |
| Conventional scanning system | Oxygenated hemoglobin | 0.936 |  |  | 0.905 |  |  |
| | Deoxygenated hemoglobin |  | 0.985 |  |  | 0.978 |  |
| | Oxygen saturation |  |  | 0.998 |  |  | 0.997 |

The maximum value of structural similarity index is 1.0 if two images are identical.

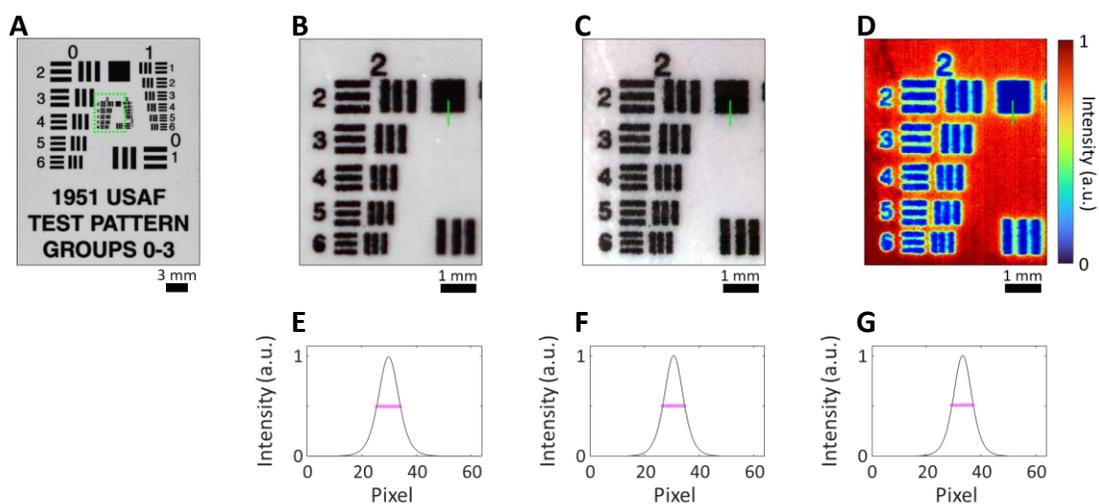

**Figure S1. Spatial resolution analyses of the dual-channel imaging setup coupled with a smartphone camera. (A)** Test target pattern used to calculate spatial resolutions. For quantitative assessments, a representative image area is marked with a green box. **(B)** Top: RGB image of the imaging area (green box in **A**) acquired by a smartphone camera. Bottom: A quantitative analysis of spatial resolution is conducted using the knife-edge method. The crossing line between the black and white areas of the edge (marked in the green line) offers an edge spread function (ESF). A line spread function (LSF) is obtained by taking the derivative of ESF with respect to the coordinate. The spatial resolution is determined by a full width at half maximum (FWHM) of LSF (magenta line), returning 9 pixels (62 µm). **(C)** Top: RGB image of the imaging area (green box in a) acquired by the RGB image camera in the dual-channel imaging setup. The RGB values in the line-scan area (slit area) are selected and restacked to form an image in a manner similar to a pushbroom-type hyperspectral imaging system. Bottom: The spatial resolution is determined by FWHM of LSF (magenta line), returning 8 pixels (55 µm). **(D)** Top: Intensity image of the imaging area (green box in **A**) at 550 nm from the hyperspectral image data (hypercube) acquired by the pushbroom-type hyperspectral imaging method. Bottom: The spatial resolution is determined by FWHM of LSF (magenta line), returning 8 pixels (55 µm).

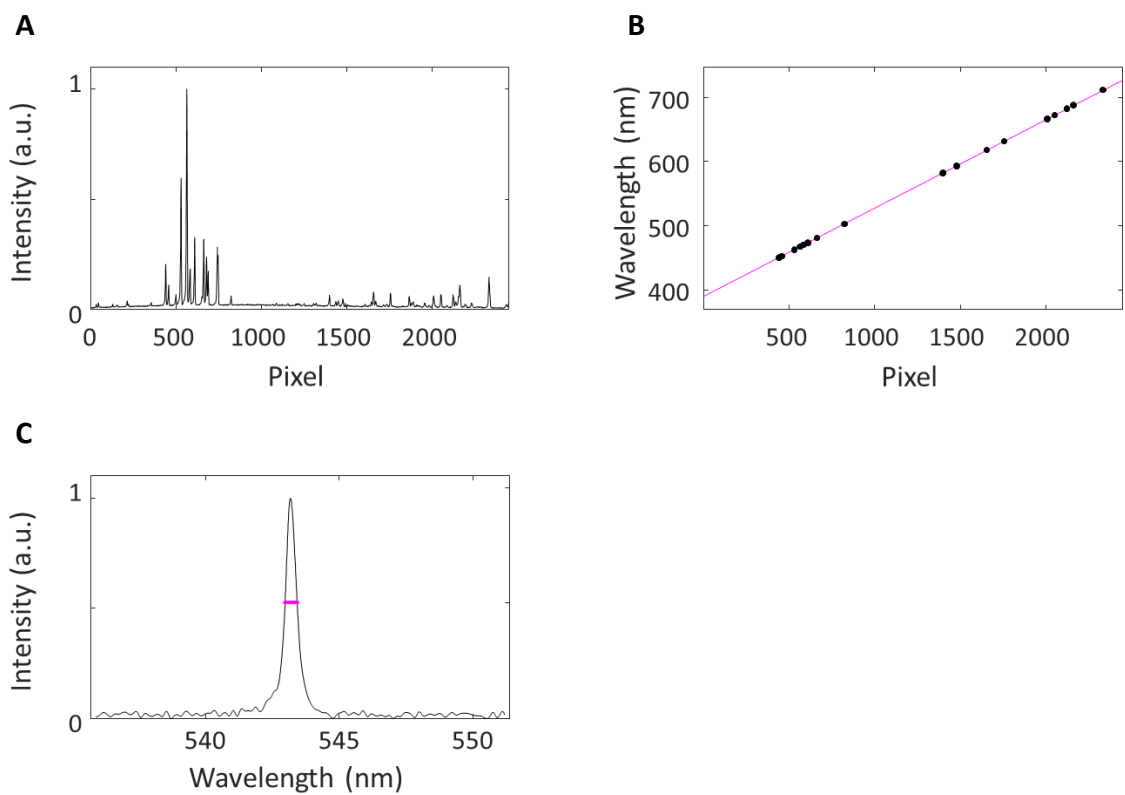

**Figure S2. Wavelength calibration and spectral resolution calculation of the dual-channel imaging setup. (A)** Xenon spectral calibration lamp (6033 Xenon Lamp, Newport) as a function of pixels. This calibration lamp produces known multiple narrow emission peaks at specific wavelengths. **(B)** Linear regression line (magenta line) used to convert the pixel number to the corresponding wavelength. The black dots are the peaks selected in **A**. **(C)** Spectral resolution Δλ determined by FWHM of the spectral profile of a HeNe laser. FWHM of the laser peak (magenta line) returns a spectral resolution Δλ of 0.5 nm and the spectral interval for hyperspectral learning is set to be 1 nm.

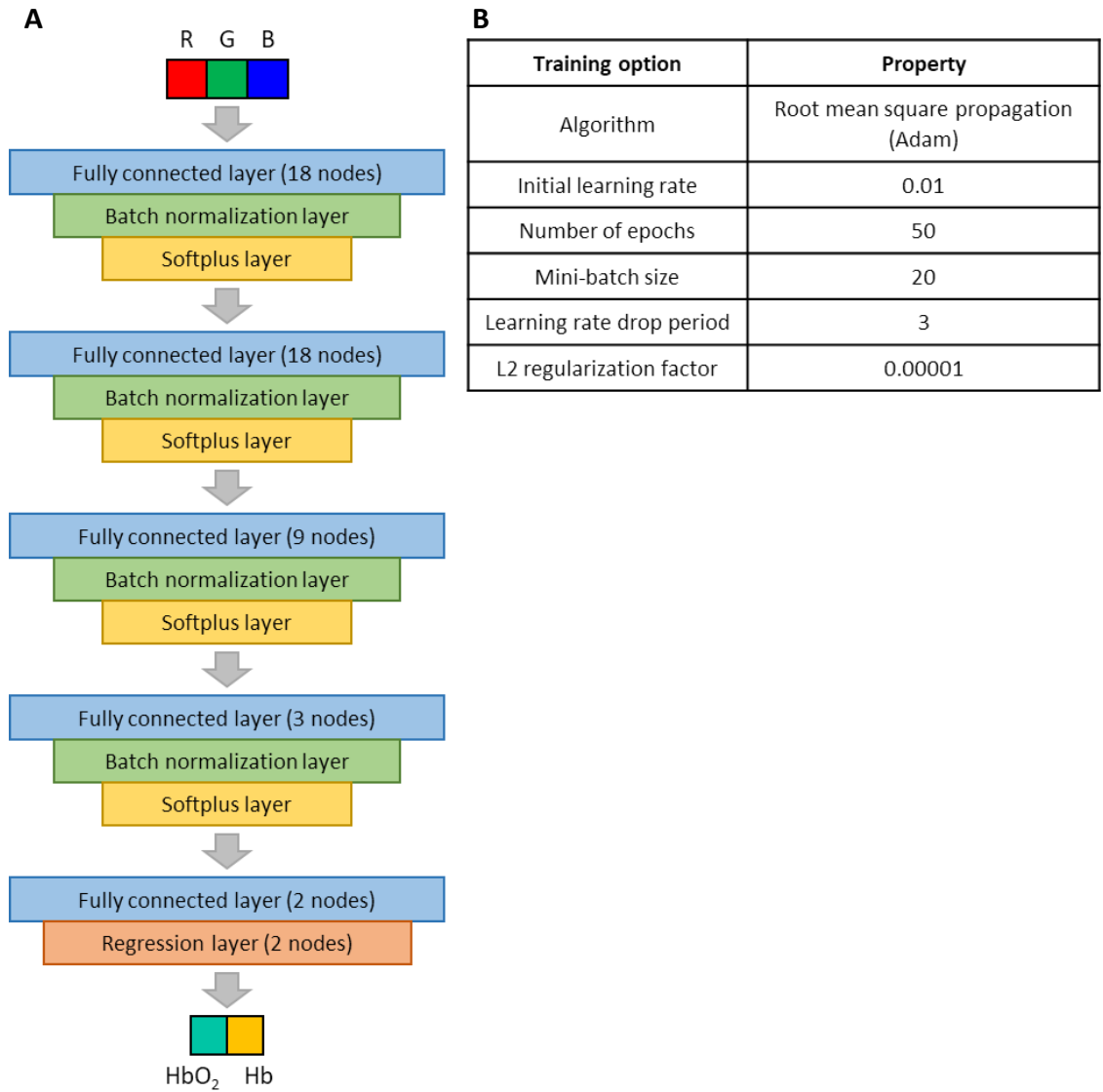

**Figure S3. Architecture diagram and key hyperparameters of the deep neural network that directly returns the hemodynamic parameters of Hb and HbO₂ from the RGB input values.** **(A)** Network is informed by hyperspectral learning as well as tissue reflectance spectral modeling such that the first hidden layer in the deep neural network is designed to mimic hyperspectral learning and the output hemodynamic parameters are extracted from tissue reflectance spectral modeling. **(B)** Training options and hyperparameters are optimized to efficiently extract the hemodynamic parameters directly from the RGB values.

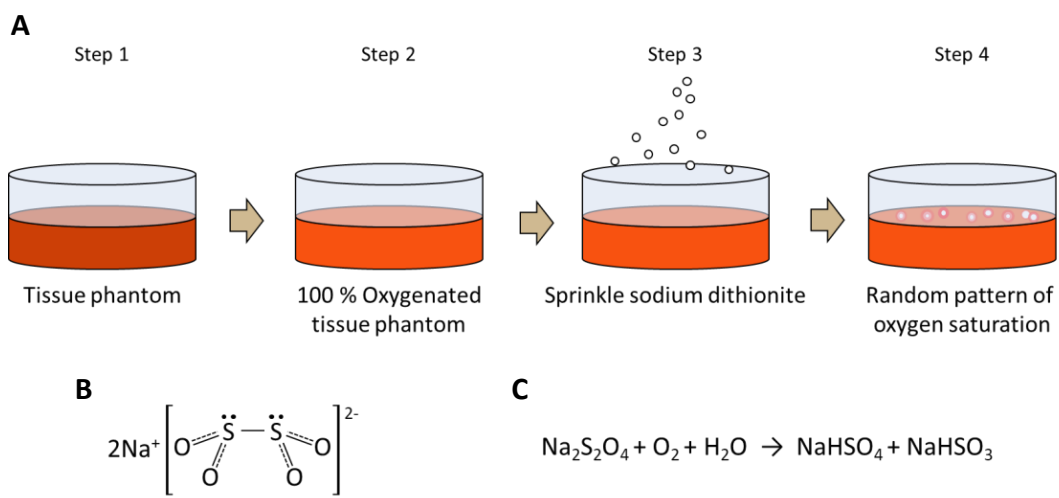

**Figure S4. Schematic illustration of tissue-mimicking phantoms that can vary oxygen saturation (sPO$_2$) and have a rapid oxygen depletion process. (A)** Tissue phantom is composed of lyophilized hemoglobin and polystyrene microspheres to mimic the light absorption and scattering properties of biological tissue, respectively. Hemoglobin is fully saturated with oxygen under standard temperature and atmospheric pressure. Sodium dithionite (Na$_2$S$_2$O$_4$) is introduced to deprive oxygen of hemoglobin. Specifically, upon sprinkling sodium dithionite, a random dynamic pattern of oxygen depletion is generated (Supplementary Methods). **(B)** Sodium dithionite. The lone electron pairs of sulfur (S) make sodium dithionite a strong reducing agent because they easily donate an electron. **(C)** Chemical equation of sodium dithionite in an aqueous solution with oxygen. In an aqueous solution (abundant with water [H$_2$O]), sodium dithionite (Na$_2$S$_2$O$_4$) and oxygen (O$_2$) can react with each other. Sodium dithionite works as an electron source when it reacts with oxygen (O$_2$) and water (H$_2$O) to generate sodium bisulfate (NaHSO$_4$) and sodium bisulfite (NaHSO$_3$) as products.

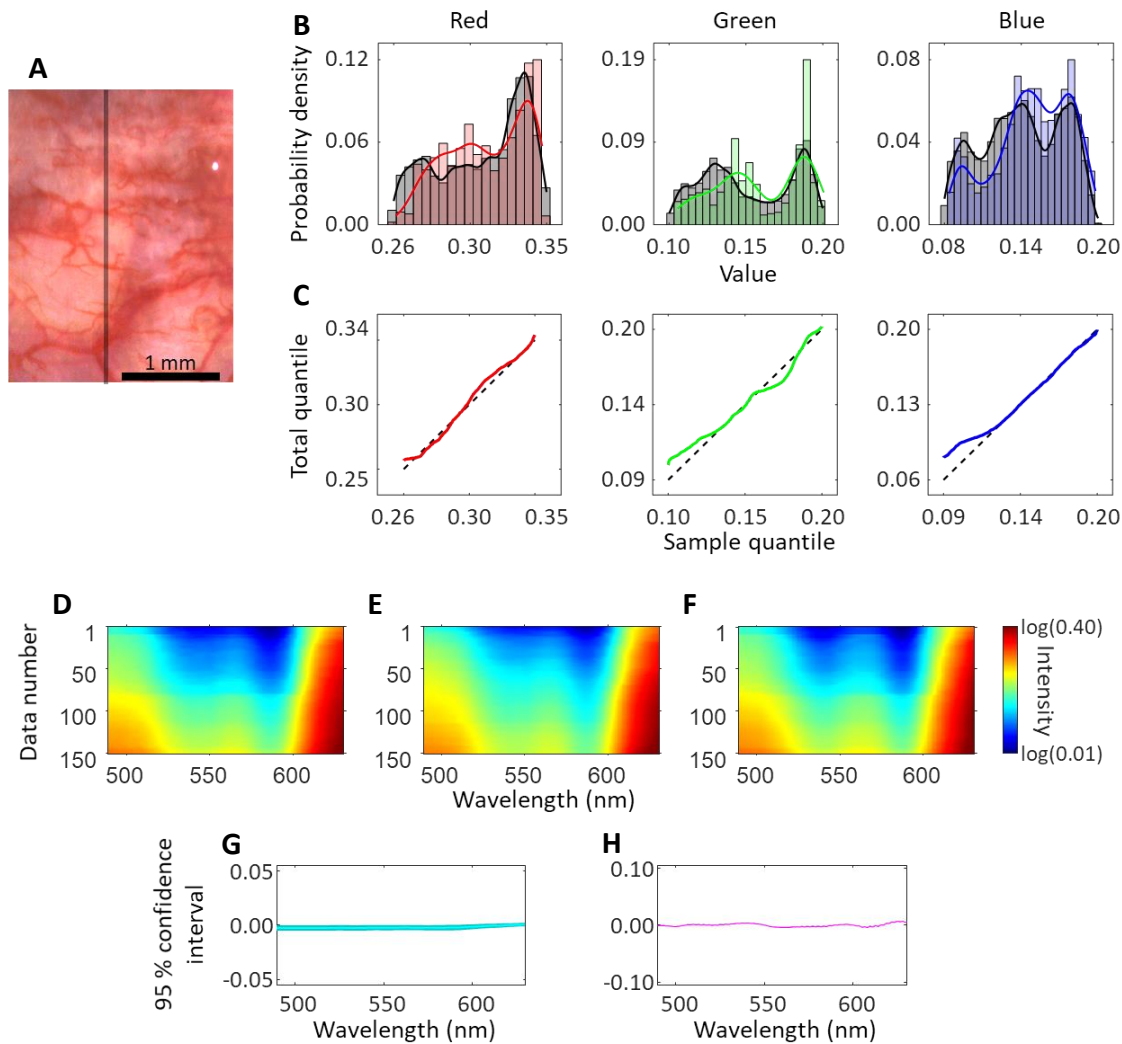

**Figure S5. Comprehensive evaluations of hyperspectral learning, hypercube recovery, and hemodynamic parameter extractions in the inner eyelid (palpebral conjunctiva). (A)** Photograph (RGB image) of the inner eyelid. The vertical gray line represents a subarea for hyperspectral sampling. **(B)** Probability distributions and histograms of R, G, and B values of the entire image area and the subarea. The black bins show the distributions of R, G, and B values of the entire image area. The black solid lines are displayed together with the black bins to visualize the trends of the data distributions of the entire image area. The colored bins depict the distributions of R, G, and B values of the subarea (i.e., the gray line in **A**) according to the RGB colors. The colored lines are illustrated together with the colored bins to visualize the trends in the data distribution of the subarea. **(C)** Quantile-quantile (Q-Q) plots for distribution comparisons between the entire field-of-view and the subarea. The subarea (gray line in a) ensures a similar intensity distribution as the intensity distributions of the entire area (**A**) in each RGB channel. The black dotted straight diagonal line indicates that the two distributions are identical. **(D-F)** Visualization of the ground-truth (**D**), reconstructed (**E**), and fitted (**F**) spectra from the testing data (data size = 150) of the sampled data (data size = 750). A logarithmic scale of intensity is used for enhanced visualization. **(G)** Comparison between the ground-truth and reconstructed spectra in the testing dataset. The differences (residuals) between the ground-truth and reconstructed spectra are plotted with 95% confidence intervals for each λ. **(H)** Comparison between the ground-truth and fitted spectra from the tissue reflectance spectral model in the testing dataset.

Hemodynamic parameters are extracted using the tissue reflectance spectral model (Methods). The differences (residuals) between the ground-truth and fitted spectra are plotted with 95% confidence intervals. The reconstructed and fitted spectra are in excellent agreement with the ground-truth spectra, supporting the performance of hyperspectral learning and tissue reflectance spectral modeling.

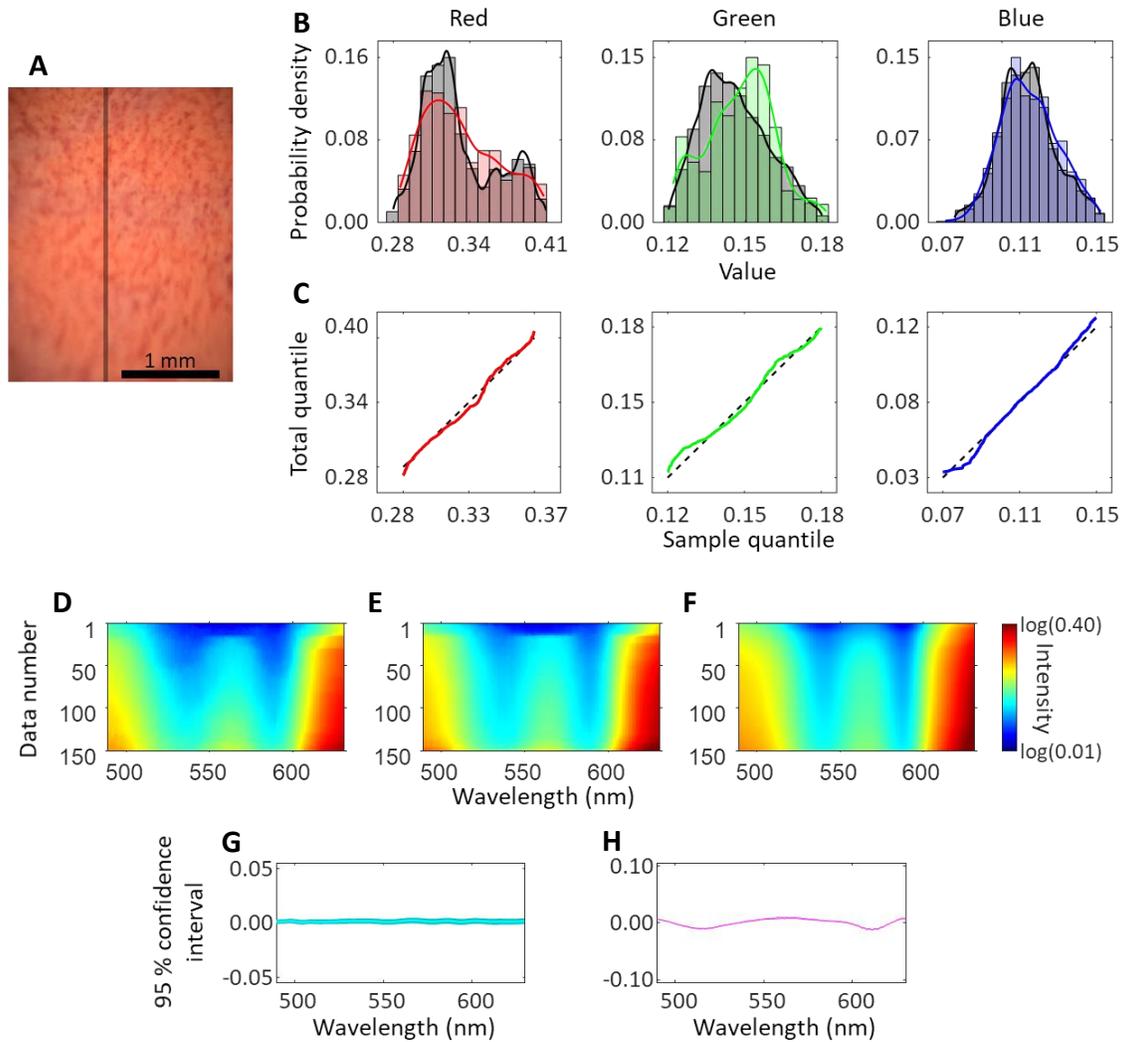

**Figure S6. Comprehensive evaluations of hyperspectral learning, hypercube recovery, and hemodynamic parameter extractions in a tissue phantom. (A)** Photograph (RGB image) of the tissue phantom having a random pattern of oxygen saturation. The vertical gray line represents a subarea for hyperspectral sampling. **(B)** Probability distributions and histograms of R, G, and B values of the entire image area and the subarea. The black bins show the distributions of R, G, and B values of the entire image area. The black solid lines are displayed together with the black bins to visualize the trends of the data distributions of the entire image area. The colored bins depict the distributions of R, G, and B values of the subarea (i.e., the gray line in **A**) according to the RGB colors. The colored lines are illustrated together with the colored bins to visualize the trends in the data distribution of the subarea. **(C)** Quantile-quantile (Q-Q) plots for distribution comparisons between the entire field-of-view and the subarea. The subarea (gray line in **A**) ensures a similar intensity distribution as the intensity distributions of the entire area (**A**) in each RGB channel. The black dotted straight diagonal line indicates that the two distributions are identical. **(D-F)** Visualization of the ground-truth (**D**), reconstructed (**E**), and fitted (**F**) spectra from the testing data (data size = 150) of the sampled data (data size = 750). A logarithmic scale of intensity is used for enhanced visualization. **(G)** Comparison between the ground-truth and reconstructed spectra in the testing dataset. The differences (residuals) between the ground-truth and reconstructed spectra are plotted with 95% confidence intervals for each λ. **(H)** Comparison between the ground-truth and fitted spectra in the testing dataset. Hemodynamic parameters are extracted using the tissue reflectance spectral model (Methods).

The differences (residuals) between the ground-truth and fitted spectra from the tissue reflectance spectral model are plotted with 95% confidence intervals. The reconstructed and fitted spectra are in excellent agreement with the ground-truth spectra, supporting the performance of hyperspectral learning and tissue reflectance spectral modeling.

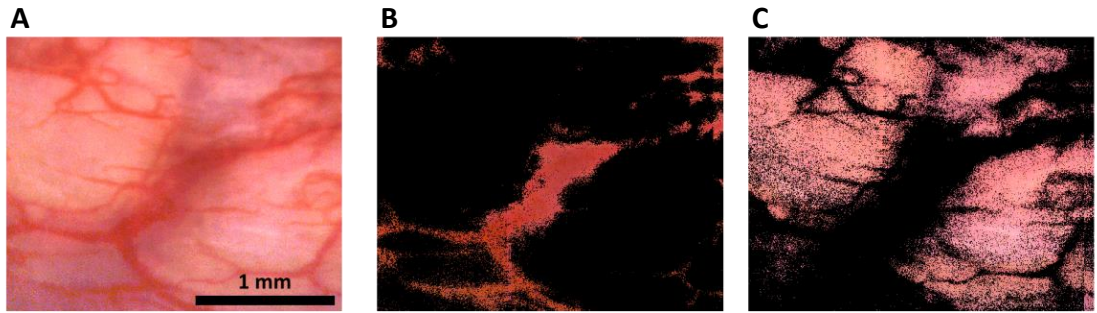

**Figure S7. Major blood vessel selection in the inner eyelid using a color threshold. (A)** Smartphone photograph (RGB image) of the inner eyelid of a healthy adult volunteer in the resting state. The major blood vessels are distinguishable by applying a set of thresholds in the RGB values. Subsequently, a binary segmentation mask is employed to select the blood vessel areas. **(B)** Selected major blood vessel areas (red areas). **(C)** Avascular tissue areas (pink bright areas).

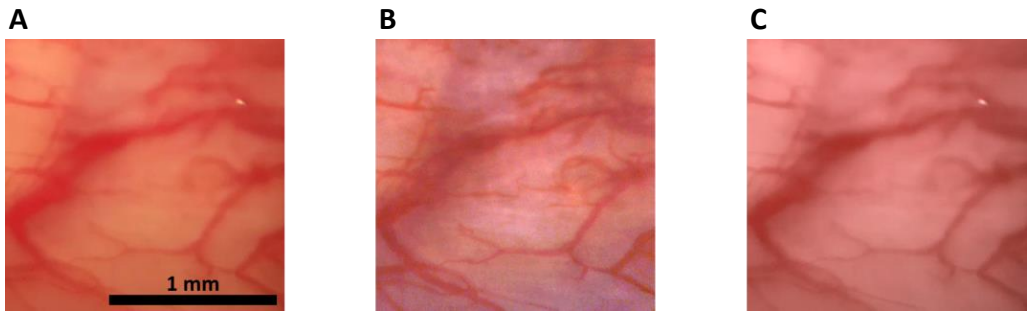

**Figure S8. Color corrections among different cameras for interoperability due to the distinct spectral responses (or sensitivity) functions. (A)** Photograph (RGB image) of the inner eyelid acquired using a smartphone camera before color correction. **(B)** RGB image of the same image area acquired using the camera in the dual-channel imaging setup. Typically, photographs obtained by digital cameras often exhibit different colors and brightness depending on the spectral response (or sensitivity) function of the camera and light conditions. **(C)** Smartphone RGB image after color correction. The relationship of RGB values between **A** and **B** can be captured by fixed-design regression, resulting in a color conversion matrix (Supplementary Methods).

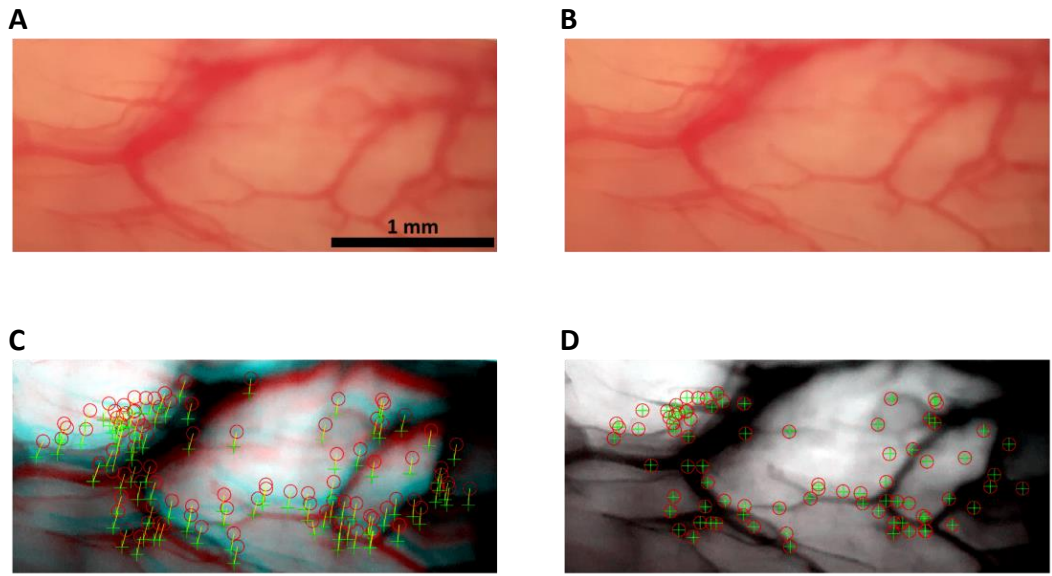

**Figure S9. Digital image stabilization of the eyelid video due to arbitrary motion artifacts (camera motion and subject motion) and optical distortions. (A)** Image frame of the smartphone video (Video S3) recording the inner eyelid. **(B)** Image frame after 60 frames (1 second) later than **A**. **(C)** Composite image showing the two neighboring image frames (**A** and **B**) overlaid in the different colors (i.e. red and cyan). When the two images are overlaid in the different colors, a slight offset and drift caused by the motion artifacts and optical distortions can be revealed. To visualize the pixel-wise differences between the two images, **A** and **B** are shown in the red and cyan colors, respectively. The salient features in **A** (green cross markers) and **B** (red circle markers) are shown on the composite image. **(D)** Overlay of the two neighboring image frames after digital image stabilization. A geometric conversion between the two frames is conducted using the set of feature points between the two different frames. The red and cyan colors are not noticeable, supporting reliable image stabilization.

**Supplementary Reference**